\documentclass[%
reprint,
superscriptaddress,
 amsmath,amssymb,
 aps,
prl,
floatfix,
]{revtex4-2}

\usepackage{graphicx}
\usepackage{dcolumn}
\usepackage{bm}
\usepackage{amsmath}
\usepackage{appendix}
\usepackage{xcolor}
\usepackage{tikz}
\usetikzlibrary{quantikz}
\usepackage{braket}
\usepackage{physics}
\usepackage[caption=false]{subfig}\usepackage{mwe}
\usepackage{graphicx}
\usepackage{blochsphere}
\usepackage{array}
\usepackage[linesnumbered,ruled,vlined]{algorithm2e}
\usepackage[thinlines]{easytable}

\usepackage{floatrow}
\usepackage{hyperref}
\hypersetup{colorlinks=true,urlcolor=blue,linkcolor=blue}

\usepackage{caption}
\newcommand{\sket}[1]{\ket{#1 \rangle}}
\newcommand{\sbra}[1]{\bra{\langle #1}}
\newcommand{\sbraket}[1]{\langle\langle #1 \rangle\rangle}

\newcommand{\superket}[1]{|{#1}\rangle\rangle}
\newcommand{\superbra}[1]{\langle\langle{#1}|}
\newcommand{\superprod}[2]{\langle\langle{#1}|{#2}\rangle\rangle}
\usepackage{multirow}

\newcommand{\AppendixName}[1]{the supplementary materials}
\newcommand{\para}[1]{}

\usepackage{cleveref}
\crefrangelabelformat{equation}{(#3#1#4--#5#2#6)}
\crefname{equation}{Eq.}{Eqs.}
\Crefname{equation}{Equation}{Equations}

\begin{document}

\preprint{APS/123-QED}

\title{Efficient characterization of qudit logical gates with gate set tomography \\using an error-free Virtual-Z-gate model
}

\author{Shuxiang Cao}
 \email{shuxiang.cao@physics.ox.ac.uk} 
 \affiliation{Department of Physics, Clarendon Laboratory, University of Oxford, OX1 3PU, UK}
\author{Deep Lall}
 \affiliation{National Physical Laboratory, Teddington, TW11 0LW, UK}
  \affiliation{Department of Materials, University of Oxford, Parks Road, Oxford, OX1 3PH, UK}
 \author{Mustafa Bakr}
 \affiliation{Department of Physics, Clarendon Laboratory, University of Oxford, OX1 3PU, UK} 
 \author{Giulio Campanaro}
 \thanks{Present address: Alice and Bob. 53 Bd du Général Martial Valin 75015 Paris, France}
 \affiliation{Department of Physics, Clarendon Laboratory, University of Oxford, OX1 3PU, UK}
 \author{Simone D Fasciati}
 \affiliation{Department of Physics, Clarendon Laboratory, University of Oxford, OX1 3PU, UK}
 \author{James Wills}
 \thanks{Present address: Oxford Quantum Circuits. Reading RG2 9LH, UK}
 \affiliation{Department of Physics, Clarendon Laboratory, University of Oxford, OX1 3PU, UK}
 \author{Vivek Chidambaram}
 \thanks{Present address: UK national quantum computing center. Didcot OX11 0QX, UK}
 \affiliation{Department of Physics, Clarendon Laboratory, University of Oxford, OX1 3PU, UK}
 \author{Boris Shteynas}
 \thanks{Present address: Oxford Quantum Circuits. Reading RG2 9LH, UK}
 \affiliation{Department of Physics, Clarendon Laboratory, University of Oxford, OX1 3PU, UK}
 \author{Ivan Rungger}
 \affiliation{National Physical Laboratory, Teddington, TW11 0LW, UK}
\author{Peter J Leek}
 \email{peter.leek@physics.ox.ac.uk} 
 \affiliation{Department of Physics, Clarendon Laboratory, University of Oxford, OX1 3PU, UK}

\begin{abstract}
Gate-set tomography (GST) characterizes the process matrix of quantum logic gates, along with measurement and state preparation errors in quantum processors. GST typically requires extensive data collection and significant computational resources for model estimation. We propose a more efficient GST approach for qudits, utilizing the qudit Hadamard and virtual Z gates to construct fiducials while assuming virtual Z gates are error-free. Our method reduces the computational costs of estimating characterization results, making GST more practical at scale. We experimentally demonstrate the applicability of this approach on a superconducting transmon qutrit.

\end{abstract}

\maketitle

\para{Introduction}
Characterizing and modeling errors is essential for the development of quantum processors. Understanding these details enables the improvements of hardware components to eliminate errors, mitigation through quantum error mitigation strategies \cite{ErrorCircuits2017,PracticalApplicationsEndo2018} and evaluating the applicability of quantum error correction codes on a quantum processor \cite{Mądzik2022,Blume-Kohout2017}. While these characterization methods were initially developed for processors based on two-level quantum systems (qubits), the recent advancement of $d$-level systems (qudits) ($d>2$) requires compatible tools for characterizing the performance of qudit logic gates. Qudit-based processors can potentially overcome technical challenges: they can reduce the number of elementary units in a physical device \cite{MultivaluedLogicGates,ExtendingQutrits,QuditsComputingWang2020,Factoringarchitectures}, improve computational efficiency \cite{ParallelismQudits,Di2012ElementaryCircuit,Li2013,QuditsComputingWang2020}, and simplify the implementation of quantum gates \cite{Lanyon2009,DecomposingQutrits}.

Randomized benchmarking (RB) is widely used to extract average gate infidelity \cite{Unitary2designRB,Emerson_2005} and has recently been demonstrated on a superconducting qutrit \cite{MorvanQutritBenchmarking,KononenkoCharacterizationBenchmarking}. While RB can determine the average gate fidelity of a specific gate, it does not provide detailed error information. Process tomography \cite{Gladden1997ProcessApplications} is a protocol for reconstructing the quantum process of a specific gate, assuming negligible state preparation and measurement (SPAM) errors, as well as Markovianity of the noise \cite{Agarwal_2024}. As an improvement on this method, gate-set tomography (GST) \cite{Greenbaum2015IntroductionTomography,Nielsen2021} takes the SPAM errors into account; GST performs process tomography for all gates in the gate set, including all elementary gates for state preparation and measurement. The quantum process can then be estimated by optimizing a numerical model that includes the process matrix and SPAM operators, using information from the entire gate set. GST has been applied to characterize quantum processes in superconducting qubits \cite{AshSakiExperimentalComputer2020,Proctor2022,White2020}, ion traps \cite{Blume-Kohout2017}, and nuclear spin qubits \cite{MaUniversalAtoms2021}. It has also been used for time-domain tracking to analyze parameter drift in quantum control \cite{WhitePerformanceGates2021,Proctor2020}. Although the techniques for qubit tomography are well-established, the development of optimal methods for designing GST experiments for qudits still requires further exploration. 

In this letter, we propose and demonstrate an efficient GST method for the characterization of qudits. Our method utilizes only qudit Hadamard gate and virtual Z gates for state preparation and measurement in different bases. By assuming that the virtual Z gates are ideal, we simplify the model to reduce the computational cost of the GST estimation process. We implement this method on a superconducting transmon qutrit, extracting the full process matrices and SPAM errors to validate its practicality. We compare the characterization results with those from a model that fully parameterizes the virtual Z gates and those assuming ideal virtual Z gates, as well as with results from RB demonstrating its validity.

The goal of GST is to reconstruct the quantum process of all the gates in the gate set \(\mathcal{G}\), taking into account that the initial state preparation and measurements are imperfect. The following discussion utilizes the superoperator formalism \cite{ClassicalAndQuantumCOmputation2002}, representing the density operator \(\rho\) and measurement operator \(E\) as a vector in Schmidt-Hilbert space, denoted by superket \(\sket{\rho}\) and superbra \(\sbra{E}\), respectively \cite{Nielsen2021} (see \AppendixName{Appendix \ref{app:ptm_gellman}} for more details). In this letter, we perform maximum likelihood GST \cite{Nielsen2021gatesettomography, Greenbaum2015IntroductionTomography}, which collects the following data from the measured probability distributions:  
\begin{equation}
    m_{ijkl} = \sbraket{E_l|F^{(m)}_i G_k F^{(p)}_j|\rho_0},
\end{equation}
where \(\sket{\rho_0}\) is the initial state and \(\sbra{E_l}\) is the measurement basis that can be directly implemented on the hardware.
The fiducials \(F^{(p)}_j\) represent the quantum processes for preparing the initial states, and \(F^{(m)}_i\) for implementing the measurement basis. Sandwiched between the fiducials, \(G_k \in \mathcal{G}\) is the quantum process in the gate set, which contains all the gates we are interested in, and the gates used to implement \(F^{(m)}_i\) and \(F^{(p)}_j\). To make the GST more accurate, \(G_k\) is usually replaced with a gate sequence that amplifies the error, known as a sequence of germs \cite{Nielsen_2020}. The estimated quantum processes \(\Tilde{G}_k\) for all \(G_k\) can be found with the maximum likelihood method by minimizing the objective function

\begin{equation}
    \begin{aligned}
    &\mathcal{L}_m(\Tilde{E}_l, \Tilde{F}^{(m)}_i, \Tilde{G}_k, \Tilde{F}^{(p)}_j, \rho) \\ 
    =&\sum_{ijkl} (\sbraket{\Tilde{E}_l| \Tilde{F}^{(m)}_i \Tilde{G}_k \Tilde{F}^{(p)}_j|\rho} - m_{ijkl})^2,
    \end{aligned}
\end{equation}
where \(\Tilde{E}_l\), \(\Tilde{F}^{(m)}_i\), \(\Tilde{G}_k\), \(\Tilde{F}^{(p)}_j\), and \(\Tilde{\rho}_0\) are the estimated values for the physical operators \(E_l\), \(F^{(m)}_i\), \(G_k\), \(F^{(p)}_j\), and \(\rho_0\). Minimizing this objective function is an optimization problem that is subject to specific physical constraints, which ensure that all the estimated operators are physical~\cite{Nielsen_2020}. Additionally, the optimization problem exhibits a gauge freedom, represented by:

\begin{equation}
\begin{aligned}
    &\langle\langle \Tilde{E}_l | \Tilde{F}^{(m)}_i \Tilde{G}_k \Tilde{F}^{(p)}_j | \rho \rangle\rangle \\ 
    = &\langle\langle \Tilde{E}_l | B (B^{-1} \Tilde{F}^{(m)}_i B) (B^{-1} \Tilde{G}_k B)  (B^{-1} \Tilde{F}^{(p)}_j B) B^{-1}| \rho \rangle\rangle,
\end{aligned}
\end{equation}
where \( B \) denotes the gauge matrix. Gauge optimization is essential for accurately determining the process operator, as well as the initial state and measurement operators~\cite{Nielsen_2020}.
 
\begin{figure}[t]
    \centering
    \includegraphics[width=\linewidth]{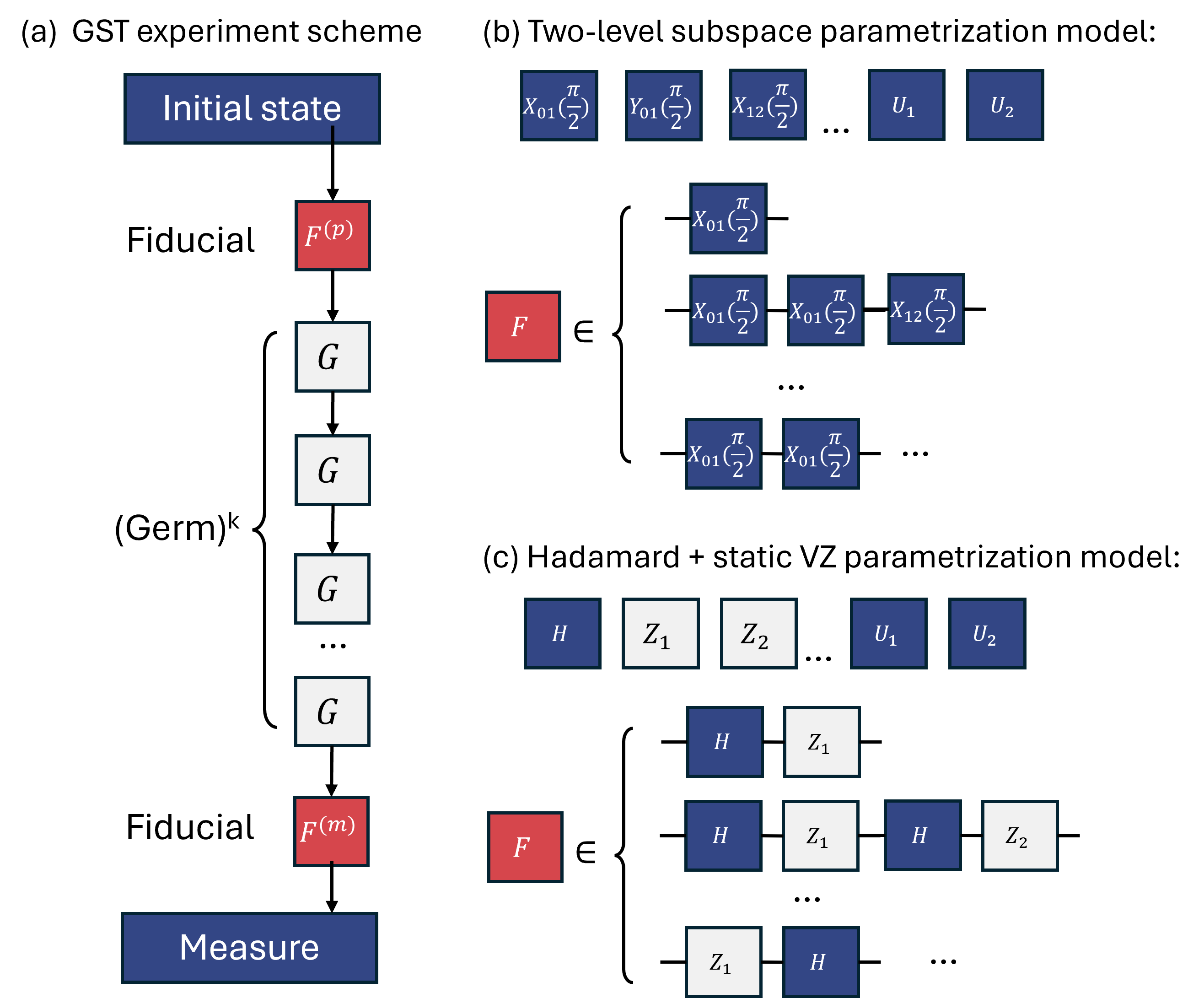}
    \caption{(a) Quantum circuit scheme for implementing GST. (b-c) Comparison between the traditional and our proposed parametrization models for GST. In these diagrams, dark-colored gates are parametrized, while light-colored gates (virtual Z gates) are fixed, reducing the number of parameters. \(U_i\) represents the gates of interest to be characterized, but not involved in constructing fiducials. Unlike the traditional approach, which parametrizes multiple gates in each neighboring two-level subspace to construct fiducials, our method requires parametrization only of the qudit Hadamard gate \(H\).\label{fig:gst_parametrization}}
\end{figure}

The above-mentioned optimizations are computationally expensive. Firstly, the number of free parameters that need to be optimized is proportional to the number of gates in the gate set used to synthesize the fiducials and germs. The optimizer for fitting the experimental data, for example, the Levenberg-Marquardt Algorithm \cite{LevenbergMarquardtAlgorithm} used by the PyGSTi software package, requires the evaluation of the Jacobian matrix at each optimization step. The size of the Jacobian matrix is proportional to the number of free parameters and to the number of elements of \(m_{ijkl}\). The computational cost of evaluating the Jacobian matrix is proportional to the length of the gate sequence~\cite{Nielsen2021}. Similar arguments apply to the Hessian matrix for error bar estimation, which has a size proportional to the square of the number of free parameters and to the number of elements in \(m_{ijkl}\) \cite{Nielsen2021} and hence even poorer scaling. Furthermore, the collection of \(m_{ijkl}\) from experiments can be time-consuming. It would therefore be best if we could minimize the size of \(m_{ijkl}\), shorten the gate sequence, and reduce the number of different gates involved in synthesizing all fiducials.

From the motivations above, we propose a new construction of the GST model optimized for cost, leveraging the existence of virtual Z gates \cite{PhysRevA.96.022330}. Unlike physical alterations of the quantum state, virtual Z gates modify the phase of subsequent gate pulses to implement a rotational frame shift, effectively applying a Z gate to the quantum state. Previous studies have shown that virtual Z gates has significantly lower errors compared to physical gates \cite{PhysRevA.96.022330}. Therefore, assuming that all virtual-Z gates are perfect can reduce the number of parameters in our model and enhance its efficiency. In addition, it also simplifies the process for gauge optimization. The gauge matrix \(B\) is no longer an arbitrary process matrix; it now has to commute with all virtual Z gates, which significantly reduces the gauge freedom. 

\begin{figure}[t]
    \centering
    \includegraphics[width=\linewidth]{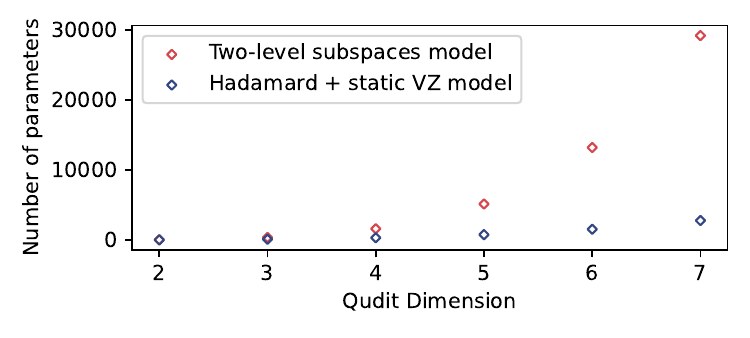}%
    \caption{The number of parameters required to parameterize the fiducial states, the initial state density operator, and the measurement operators of a single qudit, in relation to the dimension of the qudit.}
    \label{fig:parameter_number}
\end{figure}

To further reduce the model parameters, we propose constructing the fiducials using qudit Clifford gates, which can be generated by the qudit Hadamard gate \(H\) and phase gate \(S\) \cite{PhysRevA.98.032304}, defined as follows:
\begin{equation}
\begin{aligned}
    S &= \sum_{j=0}^{d-1} \omega^{j(j+1)/2} \ket{j}\bra{j},\\
    H &= \frac{1}{\sqrt{d}} \sum_{j=0}^{d-1} \sum_{k=0}^{d-1} \omega^{jk} \ket{j}\bra{k},
\end{aligned}
\end{equation}
where \(d\) is the dimension of the qudit, and \(\omega\) denotes a \(d\)-th root of unity, i.e., \(\omega^d=1\). Note that the \(S\) gate purely modifies the phase of the state and can be implemented using virtual Z gates. Therefore, the GST model for the full Clifford group can be constructed by parameterizing only the \(H\) gate, reducing the complexity of the model (see Figure \ref{fig:gst_parametrization}). Our proposed model has better scaling than the traditional approach by parameterizing gates in each neighboring two-level subspace of the qudits, and the comparison of the number of parameters required to parameterize single qudit GST fiducials is shown in Figure \ref{fig:parameter_number}. To determine the exact gates needed to construct the fiducials, we select a set of Clifford gates that produce a complete basis for preparation and measurement. These gates are selected by iteratively adding a new Clifford gate to the fiducial list, which provides access to new basis that are all orthogonal to those accessible through the already selected fiducials. With this configuration, we avoid constructing an overcomplete basis for tomography, enabling us to use the minimum number of fiducials and achieve the smallest possible size for $m_{ijkl}$. The detailed algorithms for this process can be found in \AppendixName{Algorithms \ref{algo:measurement_fiducials} and \ref{algo:preparation_fiducials} in the Appendix}.

\para{Results}

We demonstrate the experimental implementation of the proposed GST method on a superconducting transmon qutrit. This study examines the feasibility of assuming that virtual Z gates are ideal (the \textit{Static VZ model}) by comparing the results of processing the same dataset with and without parameterizing the virtual Z gates (the \textit{Full model}). The gates of interest are the $X_{01}(\pi/2)$ and $X_{12}(\pi/2)$. Additionally, we implement qutrit RB and compare its results with the outcomes of the two GST models. The qutrit characterized in this paper is a single superconducting transmon implemented in a 3D-integrated coaxial circuit design \cite{Rahamim2017Double-sided,spring2021high}. For the hardware details, please refer to \AppendixName{Appendix \ref{app:hardware_details}} and the previous study \cite{Cao_2024}. 

In this study, we choose a maximum sequence length of 512 germs, and we sample each sequence 500 times. The fiducials chosen for this experiment can be found in \AppendixName{Table \ref{tab:fiducials}}. 
We show that while previous physical proposals for tomography use an over-complete basis with 9 measurement values~\cite{ImplementationWalsh-HadamardGate,ControlandTomographyQutrit2010Bianchetti}, our proposal requires only 4 measurement values \footnote{In the experiment, we use standard dispersive measurement that can distinguish 3 different states with a single shot, determining 2 degrees of freedom of the qutrit density matrix. Each measurement obtains the probability of 3 different states; however, because they always sum up to 1, each measurement can only remove a maximum of 2 degrees of freedom. Since a qutrit density matrix has a total of 8 degrees of freedom, we require a minimum of 4 such measurements to implement qutrit state tomography.}. The collected data is processed by the PyGSTi software with our model. Overall, the Static VZ model finishes in 56.2\% of the time it takes the fully parameterized model to complete \footnote{We distribute the workload onto 20 Intel Xeon Platinum 8268 CPU cores running at 2.90GHz for both models. The fully parameterized model takes 138.5 seconds to estimate the model and 497.7 seconds to compute the Hessian for error bar estimation. The Static VZ model takes 116.1 seconds to estimate the model and 241.6 seconds to compute the Hessian.}. We expect a further increased speedup with our method as system sizes increase, because the number of parameters increases at a much slower rate with respect to the size of the system compared to the traditional GST approach.

The SPAM error is characterized by the reconstructed initial state density matrix and the measurement operators. We found the initial state infidelity $1-F_{\Tilde{\rho_0}} = 0.1137(26)$ and $0.0962(36)$ for the Full model and the Static VZ model, respectively. The average measurement infidelity is $1-F_{\Tilde{M}} = 0.0348(9)$ and $0.0307(12)$ for the Full model and the Static VZ model. For the reconstructed operators, please see \AppendixName{Figure \ref{fig:rho_and_measurement_ops}} for more details.

\begin{figure}[t]
    \centering
    \includegraphics[width=\linewidth]{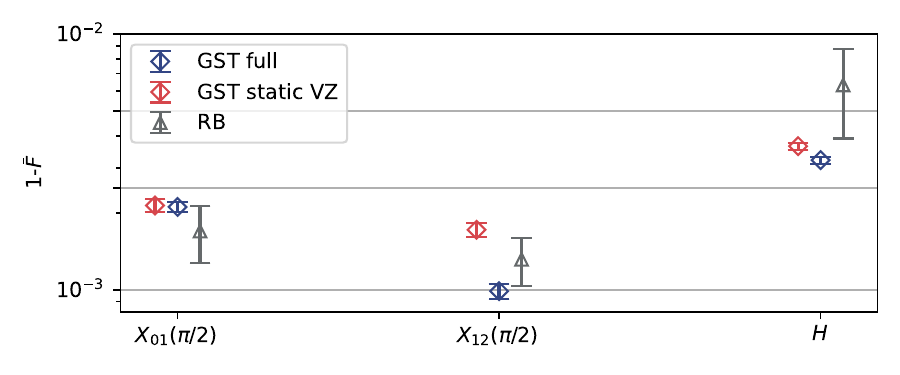}
    \caption{Comparison between GST infidelity results and the RB results. The size of the error bar indicates a 95\% confidence interval. The blue and red colors represent the GST results from two different models and the gray color represents the RB results, respectively.}
    \label{fig:Infidelity}
\end{figure}

We compare the infidelity metric between the fully parameterized GST model, the static VZ GST model, and the RB results (see Figure \ref{fig:Infidelity} and Table \ref{tab:gst_parameters}). We show that the GST results for both models obtain a comparable result to the RB results for the gates of interest. However, due to the hardware limitation of the number of gates we could implement in the RB experiment, the error bar we could obtain for the $H$ gate with RB is quite large, and we could not draw a conclusion (see \AppendixName{Appendix \ref{app:rb}} for more details).

\begin{table}[]
    \centering
\begin{ruledtabular}
\begin{tabular}{llll}
Gate            & Full $(\times 10^3)$ & Static VZ $(\times 10^3)$ & RB $(\times 10^3)$\\\hline
$I$ &  $0.365(97) $ & $0.328(96)$ & N/A\\
$Z_1(\frac{2\pi}{3})$  &  $0.221(57)$ & N/A & N/A\\
$Z_2(\frac{2\pi}{3})$  &  $0.274(70)$ & N/A & N/A\\
$X_{01}(\frac{\pi}{2})$  &  $2.112(96) $ & $2.14(12)$ & $1.70(44)$\\
$X_{12}(\frac{\pi}{2})$ &  $0.990(70) $ & $1.72(11)$& $1.32(28)$\\
$H$  & $3.22(11) $ & $3.651(12)$  & $6.3(26)$\\

\end{tabular}
\end{ruledtabular}    %
    \caption{Average gate infidelity obtained from GST experiments and RB experiments. The number in brackets in this table indicates a 95\% confidence interval.\label{tab:gst_parameters}}
\end{table}

\begin{figure}[t]
    \centering
    \includegraphics[width=\linewidth]{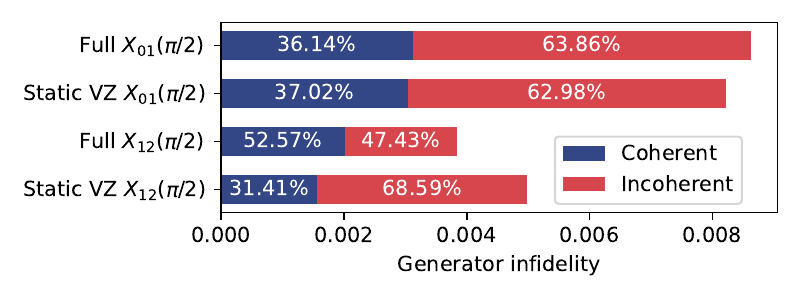}%
    \caption{ Comparison of the ratio of errors from coherent and incoherent sources as characterized by the Full model and the static VZ model. This ratio is evaluated by analyzing the contributions to the generator's infidelity from the Jamiolkowski amplitude and the Jamiolkowski probability~\cite{TaxonomyErrors}.}
    \label{fig:projection_power}
\end{figure}

\begin{figure}[t]
\subfloat[\label{fig_ptm:a} \begin{tabular}{c}
\end{tabular}]{%
  \includegraphics[width=.5\linewidth]{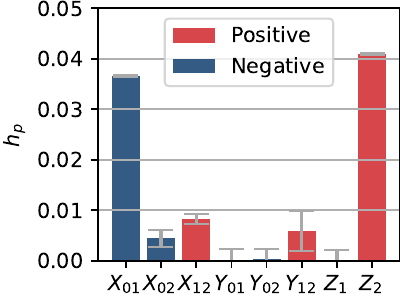}%
}
\subfloat[\label{fig_ptm:b} \begin{tabular}{c}
\end{tabular}]{%
  \includegraphics[width=.5\linewidth]{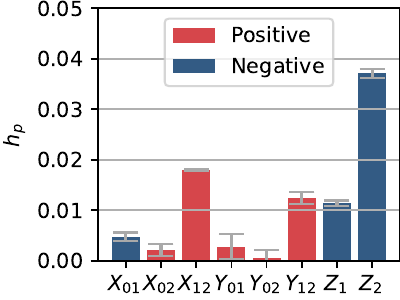}%
}\\
\subfloat[\label{fig_ptm:c} \begin{tabular}{c}
\end{tabular}]{%
  \includegraphics[width=.5\linewidth]{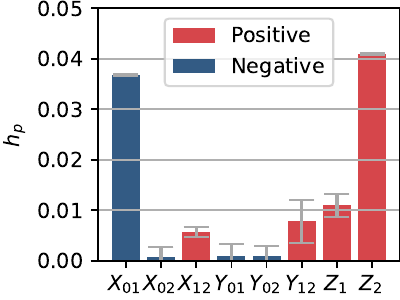}%
}
\subfloat[\label{fig_ptm:d } \begin{tabular}{c}
\end{tabular}]{%
  \includegraphics[width=.5\linewidth]{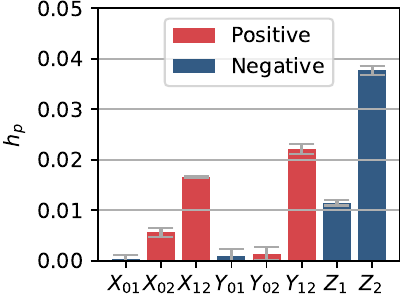}%
}
 \caption{ (a) Estimated Hamiltonian error generators for $X_{01}(\frac{\pi}{2})$ and (b) for $X_{12}(\frac{\pi}{2})$ with the full parameterization model, while (c) and (d) are obtained with the static VZ model, respectively. See \AppendixName{\ref{app:error_projection}} for more details of the projected error generators. The error bar is generated from the Hessian projection method provided in the pyGSTi package, with the confidence level set to 95\%. \label{fig:error_generator_coherence}}
\end{figure}

The advantage of GST over RB is that it provides a detailed characterization of the nature of errors. We use \textit{error generators}~\cite{TaxonomyErrors} to examine the error details. For an ideal target map $\hat{G}_i$ and a physically reconstructed map $G_i$, the post-gate error generator $L_i$ is defined as $G_i = e^{L_i}\hat{G}_i$. The error generator $L$ can be decomposed into $L = L_H + L_S + L_A + L_C$, which includes coherent ($L_H$), stochastic ($L_S$, $L_C$), and active ($L_A$) projected error generators. This decomposition allows us to quantify the amount of error using the Jamiolkowski probability $\epsilon_J = \sum_P s_P$ and the Jamiolkowski amplitude $\theta_J = \sum_P (h_P^2)^{\frac{1}{2}}$ \footnote{This is an approximation of the Jamiolkowski amplitude by assuming that contributions from the coherent errors in \(L_A\) and \(L_C\) are negligible. For further details, refer to the discussion in \cite{TaxonomyErrors}.}, where $s_P$ and $h_P$ are the coefficients when projecting $L_H$ and $L_S$ into basis $P$ (see \AppendixName{Appendix \ref{app:error_projection}}). These metrics contribute to the generator infidelity $\hat{\epsilon} = \epsilon_J + \theta_J^2$ and are used to quantify the contribution of the coherent and incoherent errors~\cite{TaxonomyErrors,Mądzik2022}. We observe that assuming a perfect virtual Z gate leads to increased incoherent errors on the $X_{12}(\pi/2)$ gate. The coherent error amounts are comparable for both models (see Figure \ref{fig:projection_power}). We further studied the details of coherent errors by analyzing the terms of the coherent error generators, as illustrated in Figure \ref{fig:error_generator_coherence}. Both models reported a similar distribution of coherent error generators, particularly in identifying the major contributors to coherent errors. While a detailed analysis of the origins of these errors is outside the scope of this study, our findings confirm that both models lead to the same conclusions regarding coherent errors. This supports the notion that our simplified model does not affect the characterization of coherent errors.

\para{Conclusion} 

In conclusion, we propose a simplified parameterization model for qudit GST that utilizes virtual Z gates, which are assumed to be perfect. Our model selects fiducials from the qudit Clifford group, allowing them to be parameterized only with the qudit Hadamard gate. This configuration simplifies the GST model and reduces the cost of applying GST on qudit processors. We conducted experimental demonstrations of our proposed GST model and a fully parameterized GST model using a superconducting transmon qutrit dataset. Our proposed method completes processing this dataset in about half of the time required by the fully parameterized model. Our findings demonstrate that the GST results from both models provide estimates of the average infidelity of the gates of interest that are close to those obtained from RB, as indicated by the 95\% confidence level. Additionally, we analyzed the Jamiolkowski amplitude and Jamiolkowski
probability and found that by assuming ideal virtual Z gates, the contribution of incoherent errors may increase. Both GST models report similar results for coherent errors, effectively capturing the major contributing terms of the coherent error generator. We conclude that our model is a reliable approach for capturing the essential errors and serves as an efficient method for qudit GST.

\begin{acknowledgments}
S.C.~acknowledges support from Schmidt Futures. P.L.~acknowledges support from the EPSRC [EP/T001062/1, EP/N015118/1, EP/M013243/1]. M.B. acknowledges support from EPSRC QT Fellowship grant EP/W027992/1. D.L. and I.R. acknowledge the support of the UK government department for Science, Innovation and Technology through the UK National Quantum Technologies Programme. We thank Heng Shen, Yutian Wen, Xianxin Guo for insightful discussions. We thank Corey Ostrove for helping resolving technical issues with pyGSTi. The authors would like to acknowledge the use of the University of Oxford Advanced Research Computing (ARC) facility in carrying out this work \cite{richards_2015_22558}.
\end{acknowledgments}


\bibliography{ref}

\appendix
\begin{widetext}

\section{Hardware details}\label{app:hardware_details}

A transmon is a multilevel quantum Duffing oscillator consisting of a Josephson junction and a shunting capacitance \cite{TransmonKoch}. The basic device parameters are given in the Table \ref{tab:device_parameters}. We utilize the first three transmon energy levels as our qutrit, labelling them as $\ket{n}$, where $n=0,1,2$. We employ as physical gates the same ones that would be used for single qubit gates on the $\{\ket{0},\ket{1}\}$ and $\{\ket{1},\ket{2}\}$ subspaces, which corresponds to driving transitions between adjacent energy levels. The qutrit virtual Z gate is included to complete a universal gate set; it is an extension of the qubit virtual Z gate~\cite{McKay2017EfficientZGates} implemented by shifting the phase of the subsequent pulses.

The tomography method requires implementing qutrit Clifford gates, which can be synthesized by qutrit Hadamard gate $H$ and virtual Z gates. $H$ is implemented with the following sequence of gates~\cite{MorvanQutritBenchmarking}:

\begin{equation}
\label{eq:hadamard_decompose}
H = H_{12}Z_{1}(\pi)Z_{2}(\pi)Y_{01}(\theta_{\text{m}})H_{12},
\end{equation}
where~$\theta_{\text{m}} = 2 \arccos(1/\sqrt{3})$ is the magic angle \cite{TheMagicAngle}, and $Y_{01}({\theta_{\text{m}}})$ is implemented with a single pulse modulating the amplitude. The virtual gates $Z_1(\theta)$ and $Z_2(\theta)$ add phases on the $\ket{1}$ and $\ket{2}$ states, respectively. $H_{12}$ is the Hadamard gate in the $\{\ket{1},\ket{2}\}$ subspace, which is synthesised by $H_{12} = Y_{12}(\pi/2) Z_2(\pi)$, where $Y_{ij}(\theta)$ is the Pauli $Y$ rotation with rotation angle $\theta$ applied to the $\{\ket{i}, \ket{j}\}$ subspace.

The basic characterization result of the device is shown in Table \ref{tab:device_parameters}. The physical single qutrit gate is implemented with 30 nanoseconds microwave pulses with DRAG correction \cite{DRAG2009} and Blackman window.  The control pulses RF signal is created by mixing with a IF signal generated by a 2 Gsps DAC and a fixed LO signal. To cover the transition frequency of the $\{\ket{0},\ket{1}\}$, $\{\ket{1},\ket{2}\}$ and $\{\ket{2},\ket{3}\}$, the LO frequency is chosen to be 3.904 GHz, which causes the IF frequency of $\{\ket{0},\ket{1}\}$ transition to be 230 MHz. 230 MHz is at the very edge of our DAC's operatable frequency range. We suggest that this can be the cause  of the finding that the gates in the $\{\ket{0},\ket{1}\}$ subspace have lower fidelities than those in the $\{\ket{1},\ket{2}\}$ subspace.

\begin{table}[]
    \centering
      \begin{ruledtabular}
        \begin{tabular}{crlc}
        Subspace&Parameter&& Value\\\colrule 
        & Resonator frequency &$f_{Res}$ (MHz)& 8783   \\
            $\ket{0}, \ket{1}$ & Transition frequency &$f_{01}$ (MHz) & 4134.33  \\
            $\ket{0}, \ket{1}$ & Relaxation time  &$T_1^{(01)}$ (us)  & $221 \pm 30$ \\
            $\ket{0}, \ket{1}$ & Hahn decoherence time &$T_2^{(01)}$ Echo (us) & $ 126 \pm 15$  \\
            $\ket{0}, \ket{1}$ & Ramsey decoherence time &$T_2^{(01)}$ Ramsey (us) & $ 96 \pm 10 $  \\
            $\ket{1}, \ket{2}$& Transition frequency &$f_{12}$ (MHz) & 3937.66  \\
            $\ket{1}, \ket{2}$& Relaxation time &$T_1^{(12)}$ (us)      & $119 \pm 20$  \\
            $\ket{1}, \ket{2}$& Hahn decoherence time &$T_2^{(12)}$ Echo (us) & $76 \pm 27$ \\
            $\ket{1}, \ket{2}$& Ramsey decoherence time &$T_2^{(12)}$ Ramsey (us) & $52 \pm 4 $ \\
        \end{tabular}
        \end{ruledtabular}
    
    \caption{Summary of device parameters \label{tab:device_parameters}}
\end{table}

\section{PTM and Gellman matrices\label{app:ptm_gellman}}

The superoperator formalism represents a density matrix $\rho$ as a vector (superket) $\superket{\rho} = \sum_m \superket{m} \superprod{m}{\rho}$ in a Hilbert-Schmidt space with dimension $d^2$, where $\superket{m}$ is the unit vector along each basis and $d$ is the dimension of the quantum system. The inner product in the Hilbert-Schmidt space is defined as $\superprod{\rho_A}{\rho_B} = \text{Tr}(\rho_{A}\rho_{B}^{\dagger})/d $, where $\rho_A$ and $\rho_B$ are density operators. In this representation, quantum operations can be represented as Pauli transfer matrix (PTM) $R_{\Lambda} = \sum_{m,n} \superket{m}\superbra{m}R_{\Lambda}\superket{n}\superbra{n}$ with dimension $d^2 \times d^2$. The $ij$ element of the PTM can be calculated from, 
\begin{equation} \label{eq:PTM}
    (R_{\Lambda})_{ij}=\frac{1}{d}Tr\{P_i\Lambda(P_j)\}.
\end{equation}
The $ (R_{\Lambda})_{ij}$ denote the expectation value when the state is prepared in the $P_j$ basis, and measured in the $P_i$ basis after applying the quantum process. The PTM conveniently acts on the state like
\begin{equation}
     R_{\Lambda} \superket{\rho} = \superket{\Lambda(\rho)},
\end{equation}
where $\Lambda$ denontes the quantum channel $\Lambda(\rho) = \sum_{i=1}^{N} K_i \rho K_i^{\dagger}$ and $K_i$ are Kraus operators.
Furthermore, two quantum operations on a quantum map can just be seen as applying one PTM to the state after the other
\begin{equation}
     R_{\Lambda_1 \cdot \Lambda_2 } \superket{\rho} = R_{\Lambda_1} R_{\Lambda_2} \superket{\rho} = \superket{\Lambda_2(\Lambda_1(\rho))}.
\end{equation}

The above discussion is valid for a system with arbitrary dimensions, and any basis $\superket{m}$ can be used in the Hilbert-Schmidt space. However it is useful to define a basis that is orthonormal, hermitian, and traceless for all members in the basis apart from the identity. For the 3-level system demonstrated in this letter, we chose the Gell-Mann matrices and the qutrit identity as the basis operators, which naturally generalise the qubit Pauli basis and are orthonormal, hermitian, and traceless:

\begin{equation}
\begin{aligned}
    &I = \begin{pmatrix}
    1 & 0 & 0\\ 
    0 & 1 & 0\\ 
    0 & 0 & 1
    \end{pmatrix} \qquad
    &X_{01} = \begin{pmatrix}
    0 & 1 & 0\\ 
    1 & 0 & 0\\ 
    0 & 0 & 0
    \end{pmatrix} \qquad
    &X_{02} = \begin{pmatrix}
    0 & 0 & 1\\ 
    0 & 0 & 0\\ 
    1 & 0 & 0
    \end{pmatrix} \\\\
    &X_{12} = \begin{pmatrix}
    0 & 0 & 0\\ 
    0 & 0 & 1\\ 
    0 & 1 & 0
    \end{pmatrix} \qquad
    &Y_{01} = \begin{pmatrix}
    0 & -i & 0\\ 
    i & 0 & 0\\ 
    0 & 0 & 0
    \end{pmatrix} \qquad
    &Y_{02} = \begin{pmatrix}
    0 & 0 & -i\\ 
    0 & 0 & 0\\ 
    i & 0 & 0
    \end{pmatrix} \\\\
    &Y_{12} = \begin{pmatrix}
    0 & 0 & 0\\ 
    0 & 0 & -i\\ 
    0 & i & 0
    \end{pmatrix} \qquad
    &Z_{1} = \begin{pmatrix}
    1 & 0 & 0\\ 
    0 & -1 & 0\\ 
    0 & 0 & 0
    \end{pmatrix} \qquad
    &Z_{2} = \frac{1}{\sqrt{3}}\begin{pmatrix}
    1 & 0 & 0\\ 
    0 & 1 & 0\\ 
    0 & 0 & -2
    \end{pmatrix} \\
\end{aligned}
\label{eq:qutrit_basis}
\end{equation}

With the basis defined above, we present the reconstructed PTM and post-gate error generators for all parametrized gates in the gate set in Figure \ref{tab:all_ptm_report_full} and  Figure \ref{tab:all_ptm_report_full} for the Full model and the Static VZ model, respectively.

\newcommand{\iptm}[1]{\begin{tabular}{c} \includegraphics[width=4.5cm,height=4.5cm]{#1}\end{tabular}}

\begin{figure}[h]
    \centering
    \begin{tabular}{m{1.8cm}|c|c|c}
         Gate name & $I$ & $Z_1(\frac{2\pi}{3})$ & $Z_2(\frac{2\pi}{3})$  \\ &&& \\ 
         \begin{tabular}{c}Average\\infidelity\end{tabular}  & $0.000365(97)$ & $0.000294(76)$ & $0.000274(70)$\\&&& \\
         \begin{tabular}{c}PTM\\\end{tabular} & \iptm{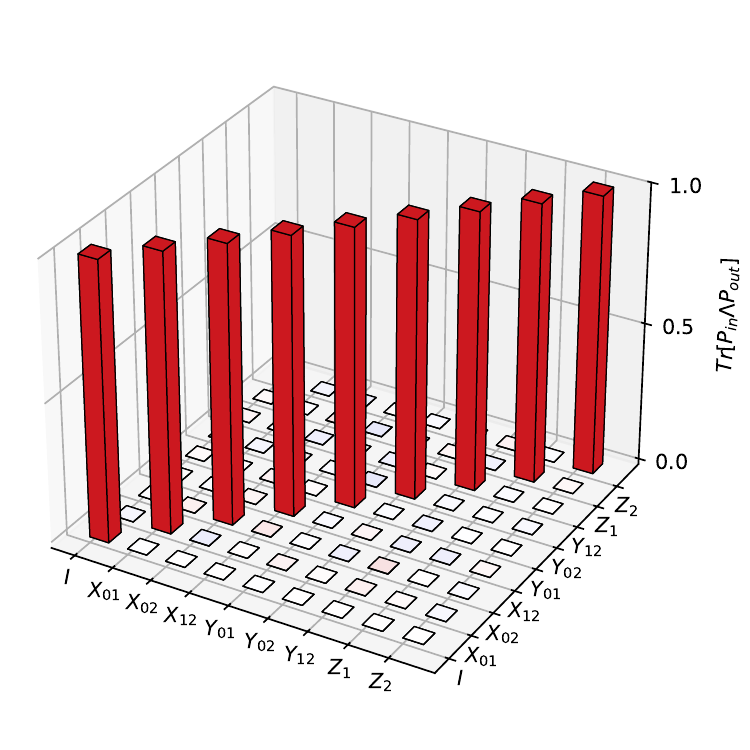} & \iptm{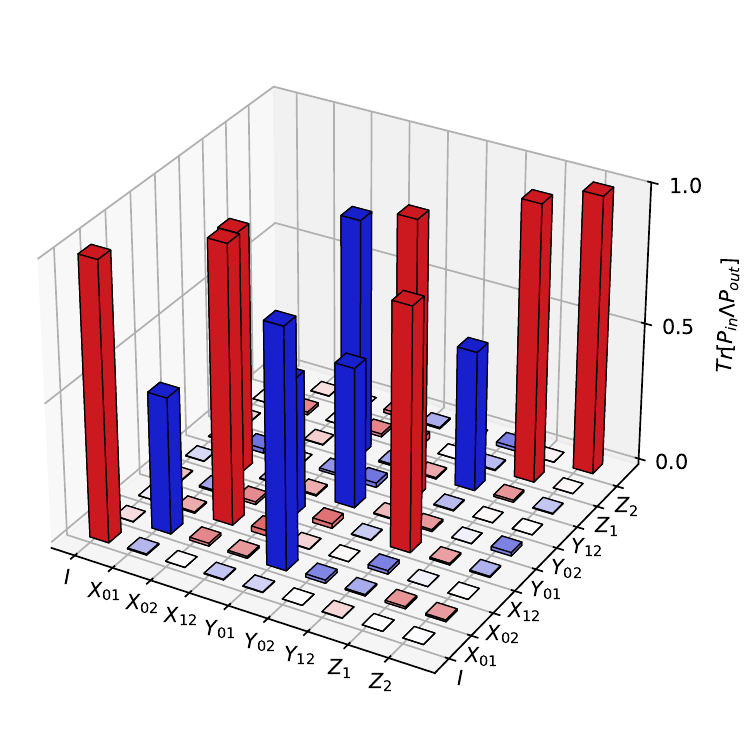} & \iptm{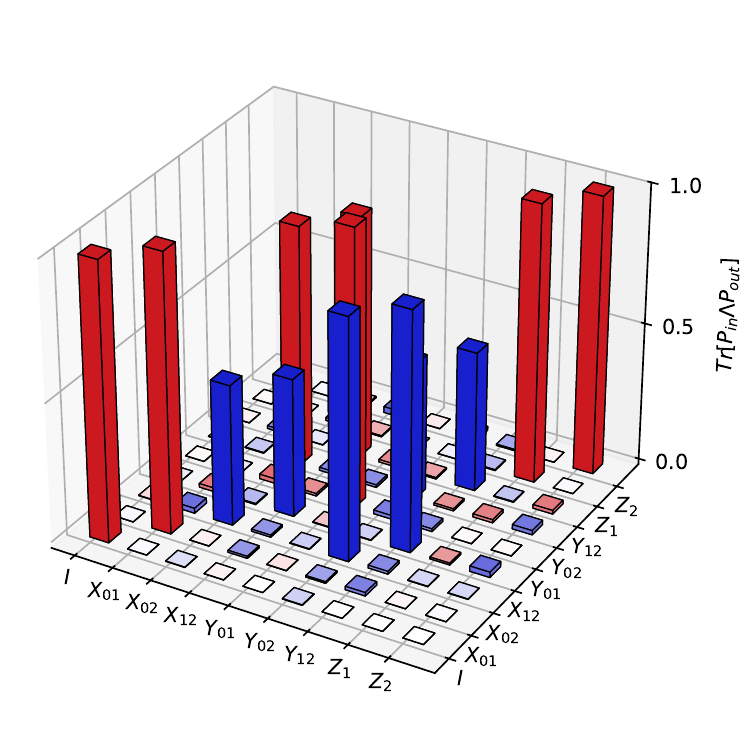} \\ &&&\\
         \begin{tabular}{c}Error\\generator\end{tabular} & \iptm{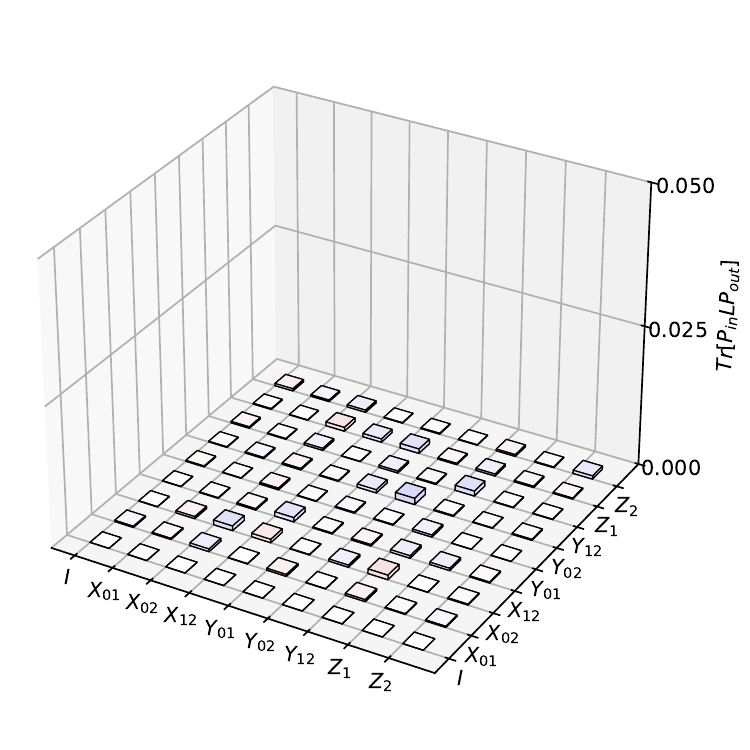} & \iptm{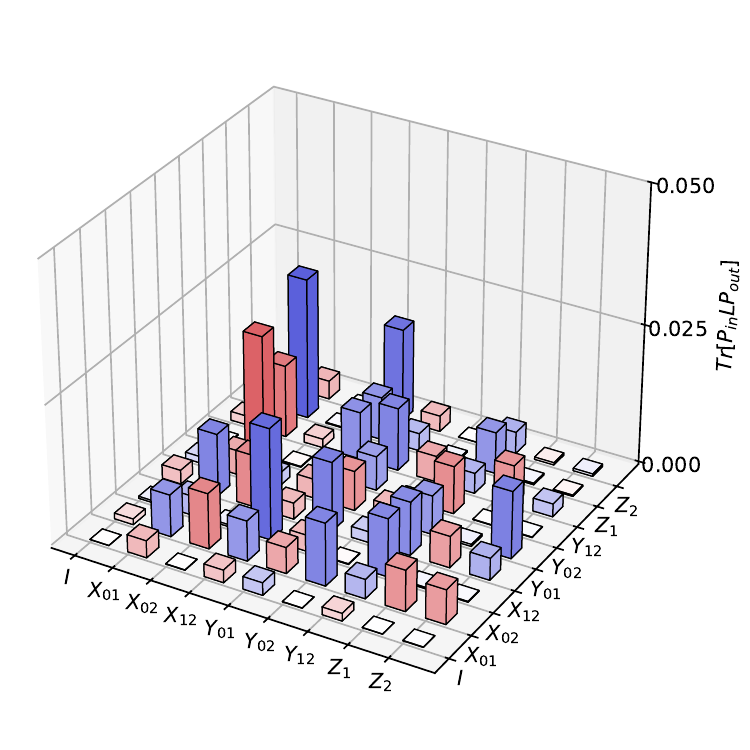} & \iptm{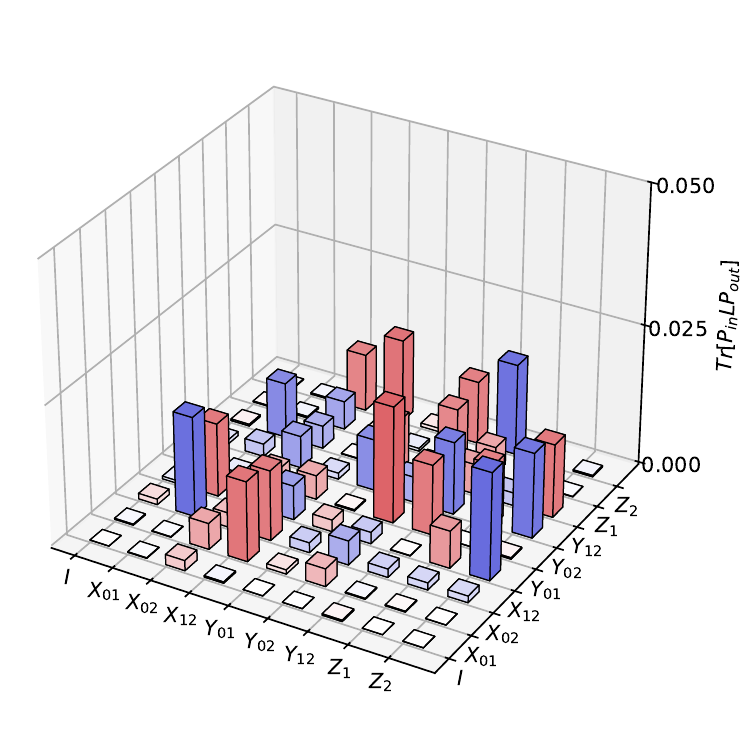}\\ \hline&&& \\
         Gate name & $X_{01}(\frac{\pi}{2})$ & $X_{12}(\frac{\pi}{2})$ & $H$  \\&&& \\
         \begin{tabular}{c}Average\\infidelity\end{tabular}  & $0.002112(96)$ & $0.00099(7)$ & $0.00322(11)$ \\&&& \\
         \begin{tabular}{c}\\PTM\\\end{tabular} & \iptm{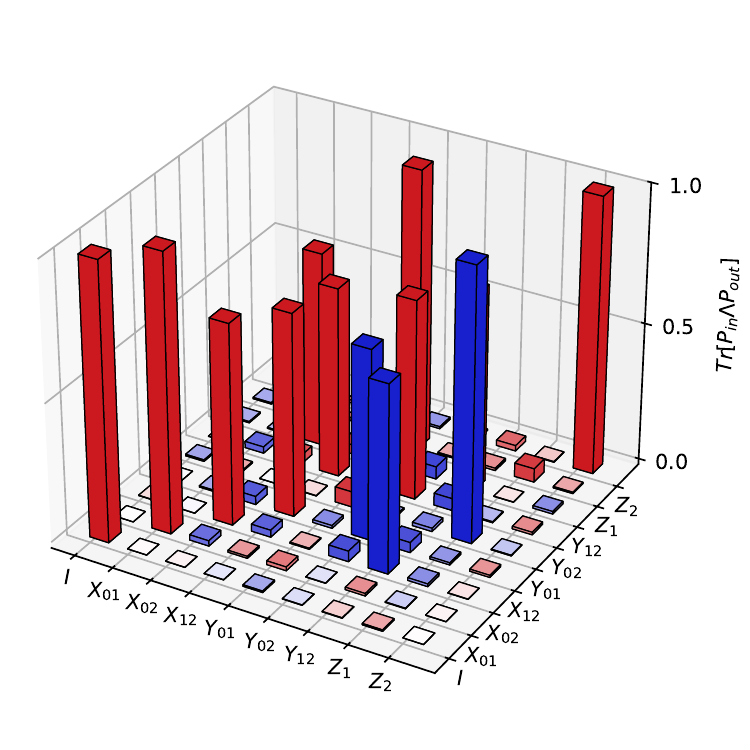} & \iptm{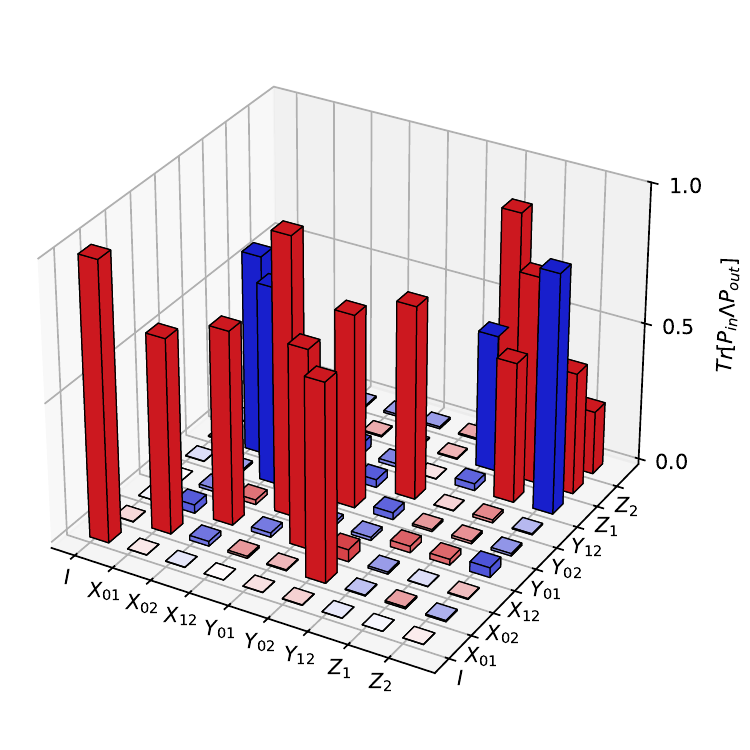} & \iptm{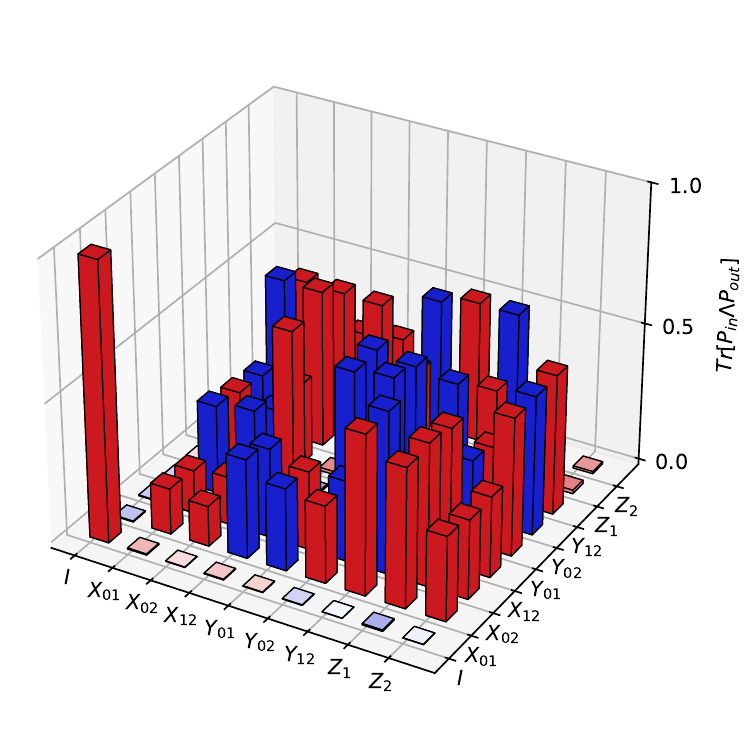} \\ &&&\\
         \begin{tabular}{c}Error\\Generator\end{tabular} & \iptm{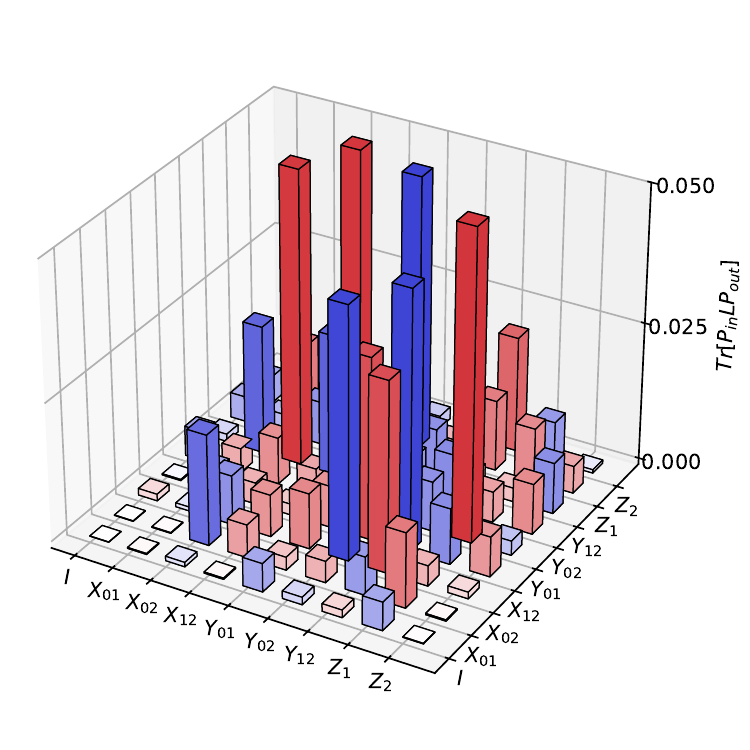} & \iptm{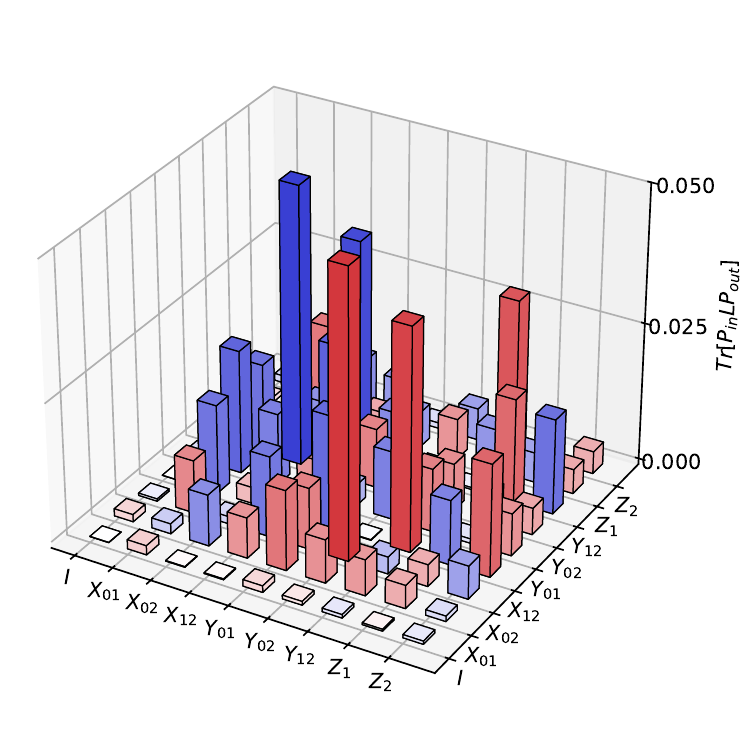} & \iptm{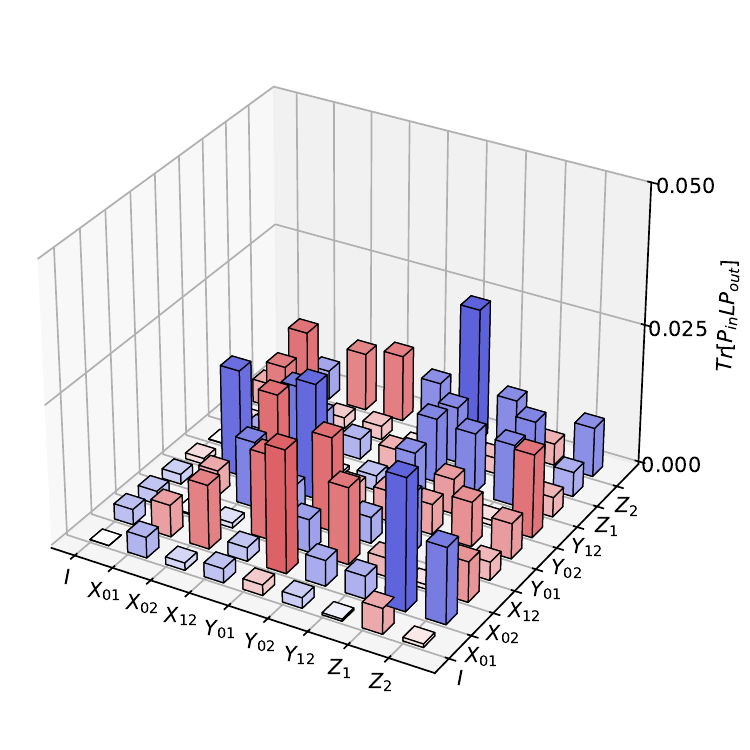}\\ 
         
    \end{tabular}
    \caption{Averaged gate infidelities, reconstructed process matrix and error generators of the gate set for full parametrization. The result is generated by pyGSTi maximum likelihood estimation.} 
    \label{tab:all_ptm_report_full}
\end{figure}

\begin{figure}[h]
    \centering
    \begin{tabular}{m{1.8cm}|c|c|c}
         Gate name & $I$ & $Z_1(\frac{2\pi}{3})$ & $Z_2(\frac{2\pi}{3})$  \\ &&& \\ 
         \begin{tabular}{c}Average\\infidelity\end{tabular}  & $0.000328(96)$ & $ N/A $ & $N/A $\\&&& \\
         \begin{tabular}{c}PTM\\\end{tabular} & \iptm{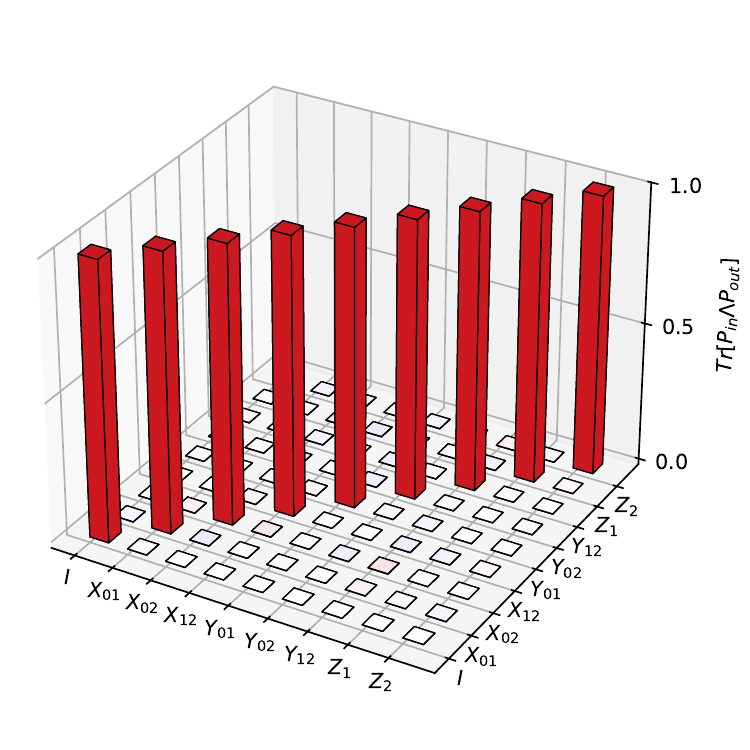} & \iptm{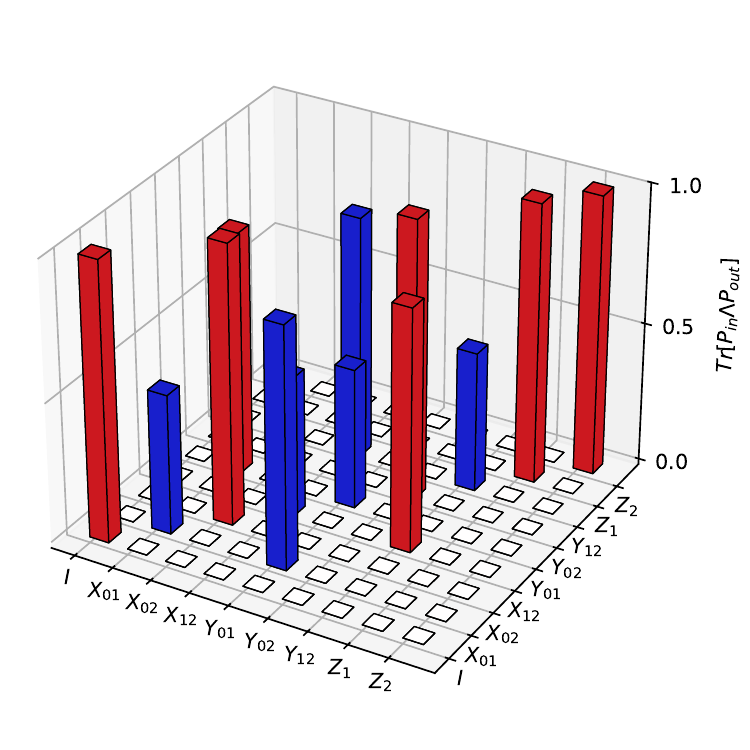} & \iptm{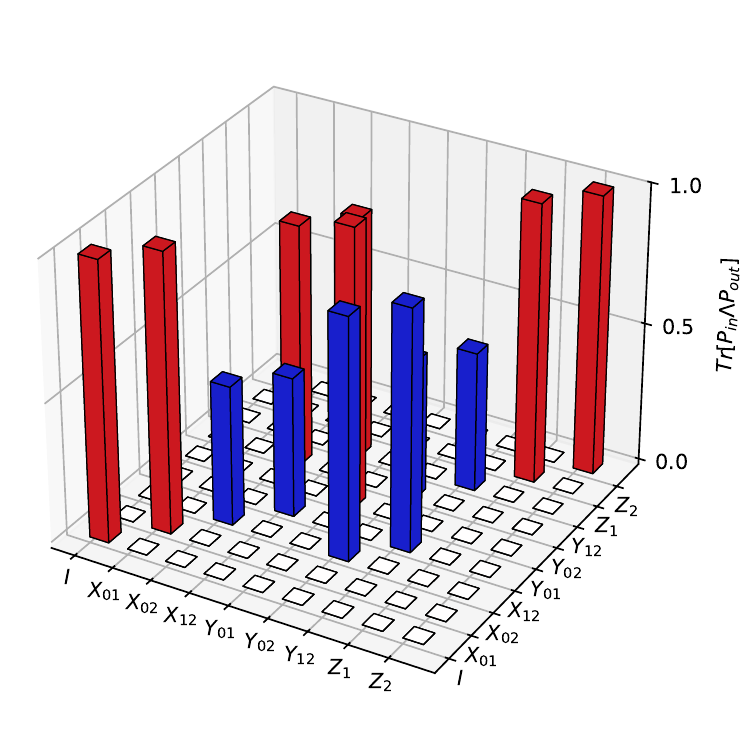} \\ &&&\\
         \begin{tabular}{c}Error\\generator\end{tabular} & \iptm{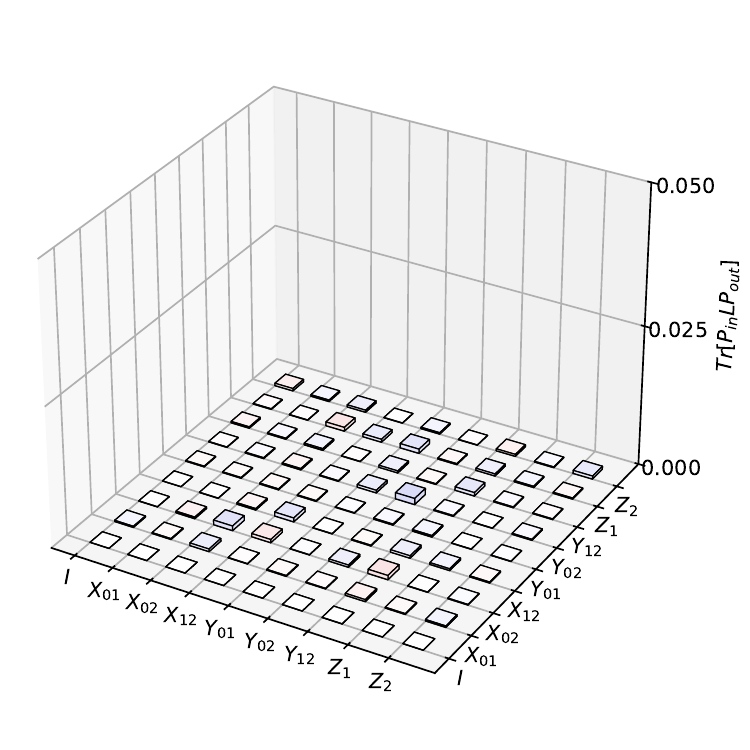} & \iptm{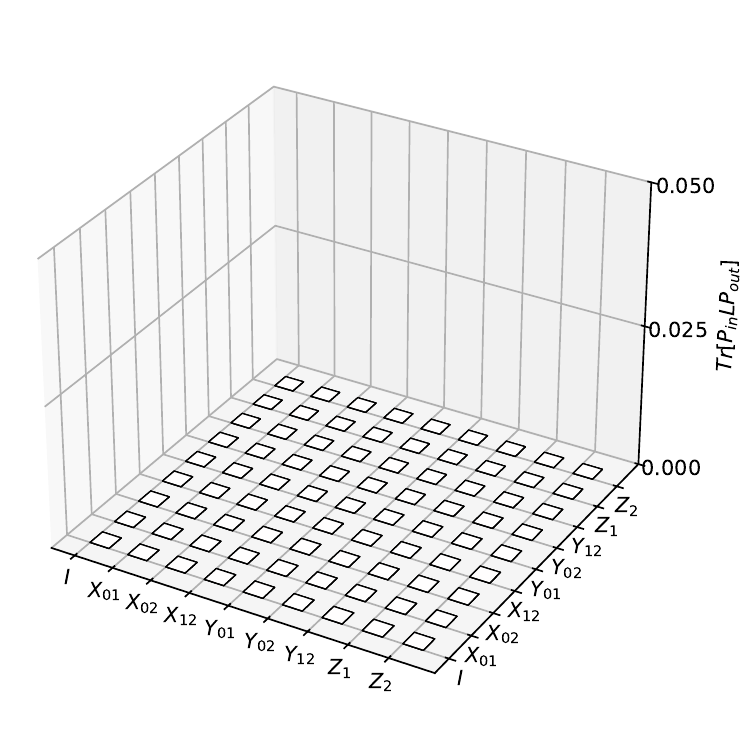} & \iptm{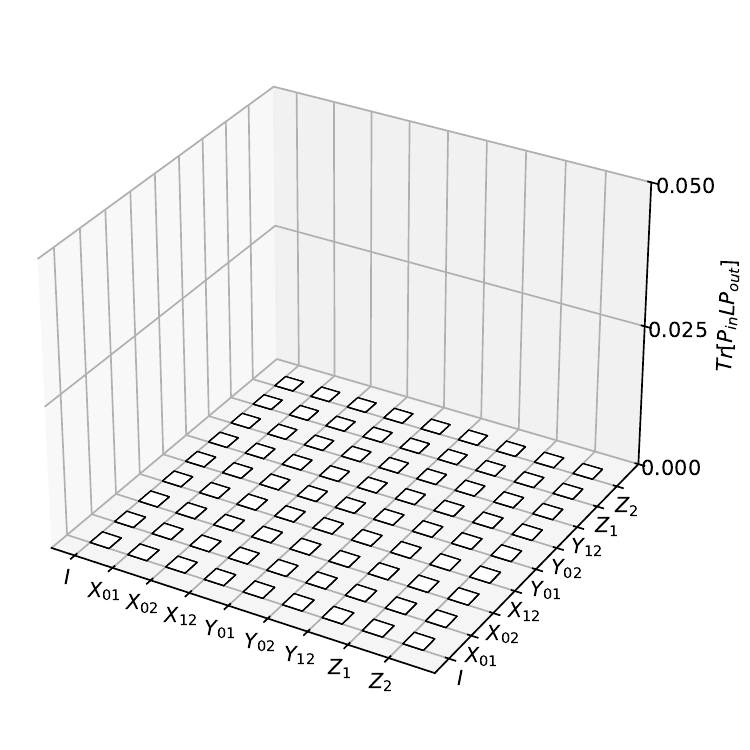}\\ \hline&&& \\
         Gate name & $X_{01}(\frac{\pi}{2})$ & $X_{12}(\frac{\pi}{2})$ & $H$  \\&&& \\
         \begin{tabular}{c}Average\\infidelity\end{tabular}  & $0.00214(12)$ & $0.00172(11)$ & $0.003651(12)$ \\&&& \\
         \begin{tabular}{c}\\PTM\\\end{tabular} & \iptm{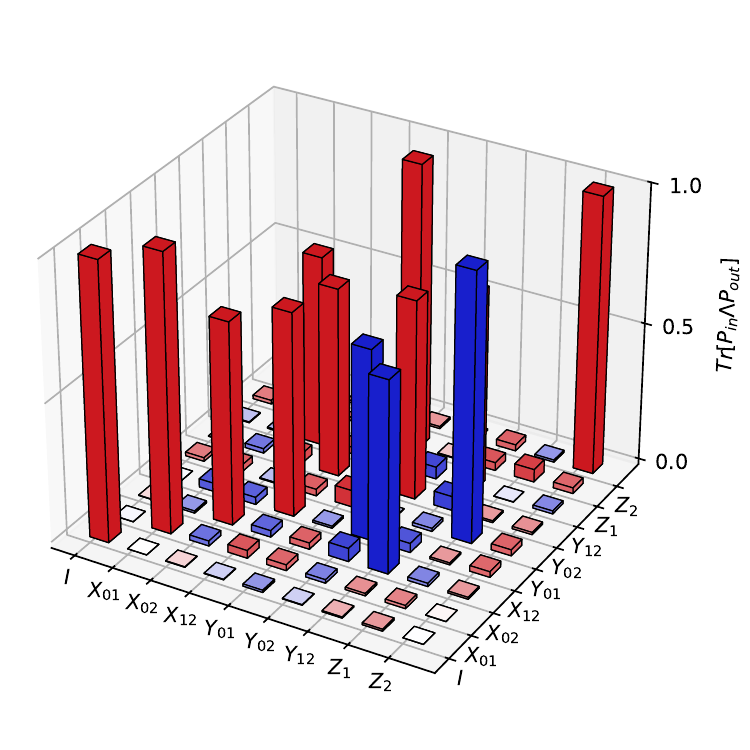} & \iptm{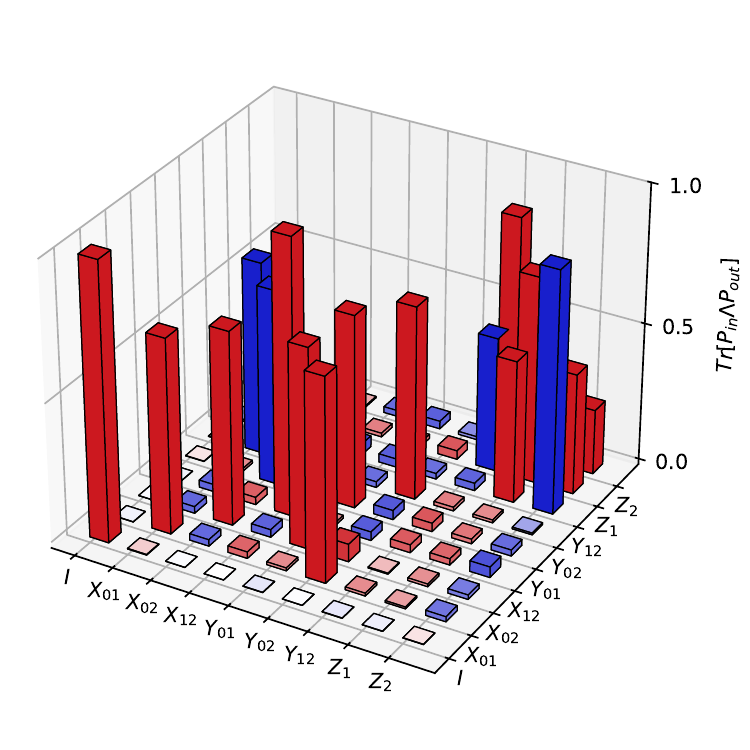} & \iptm{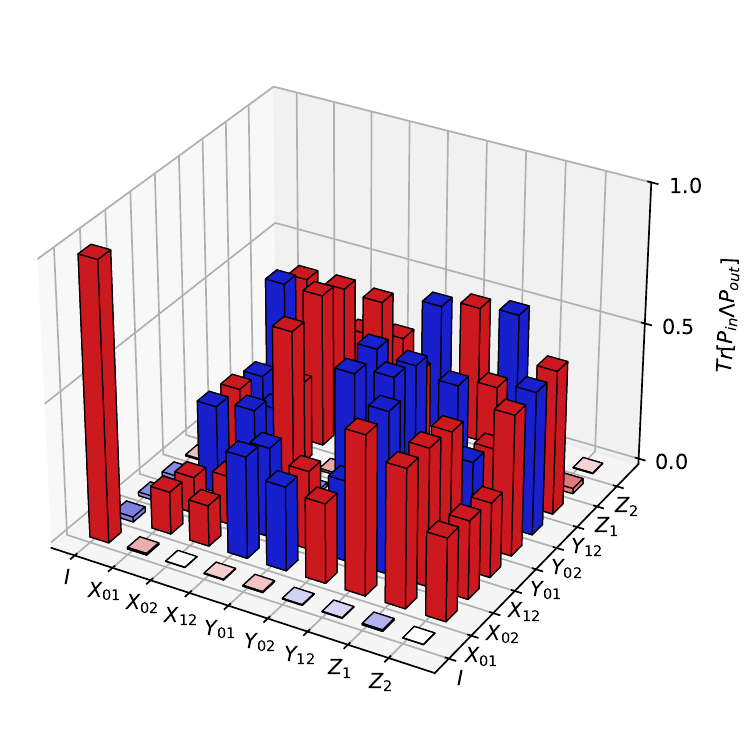} \\ &&&\\
         \begin{tabular}{c}Error\\Generator\end{tabular} & \iptm{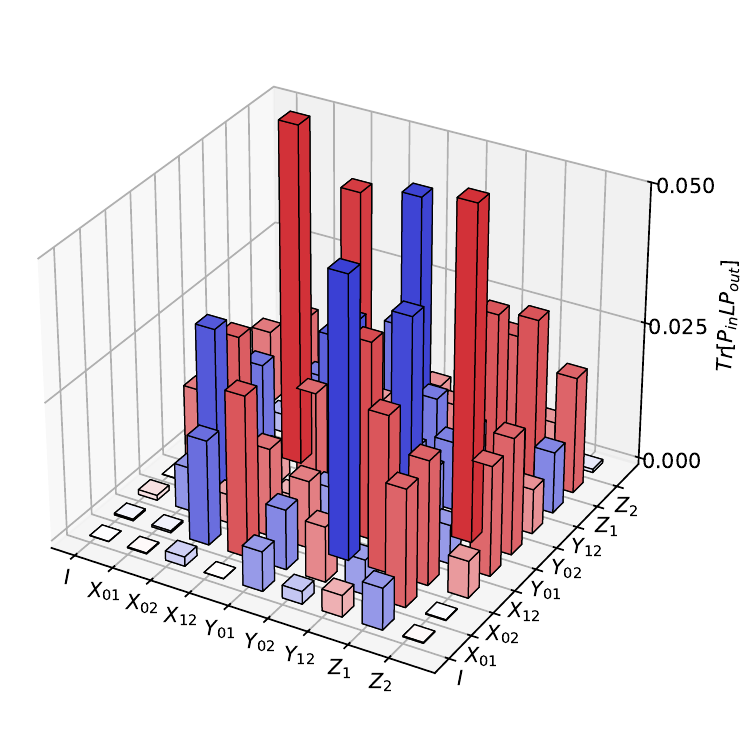} & \iptm{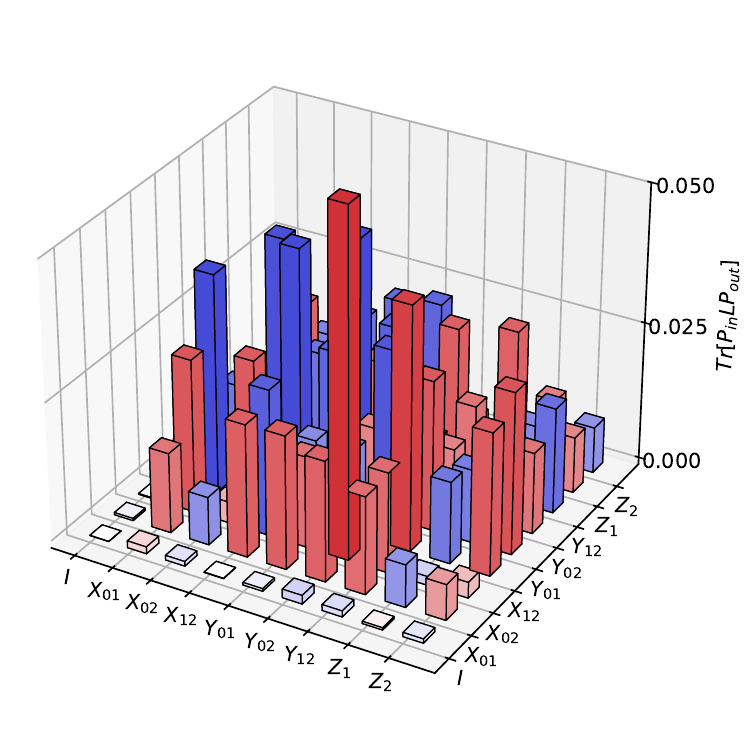} & \iptm{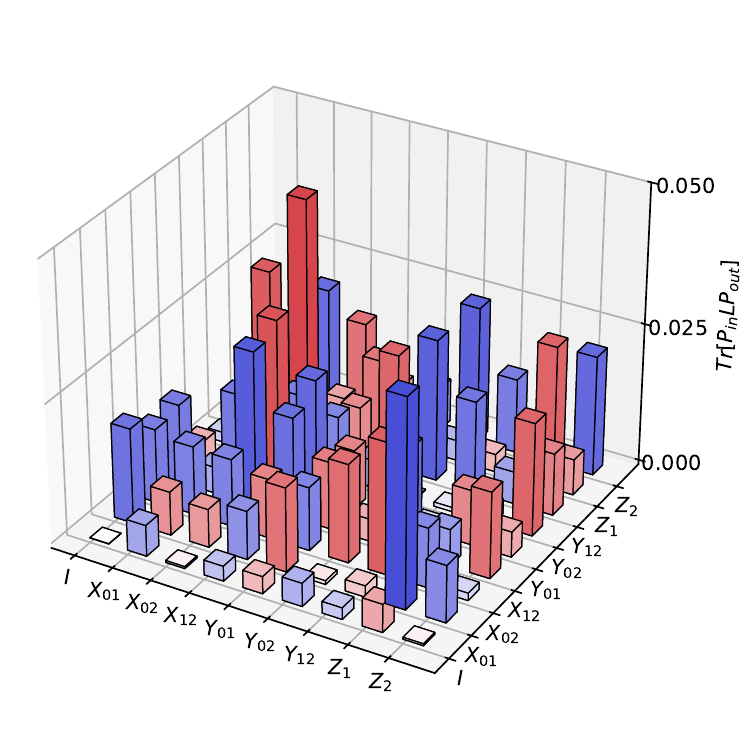}\\ 
         
    \end{tabular}
    \caption{Averaged gate infidelities, reconstructed process matrix and error generators of the gate set for static VZ parametrization. The result is generated by pyGSTi maximum likelihood estimation.} 
    \label{tab:all_ptm_report_static_Z}
\end{figure}

\begin{figure}[hbt]
\subfloat[\label{subfig:a} \begin{tabular}{c}
       Full $\Tilde{\rho_0}$\\$F_{\rho_0}=0.8863(26)$
\end{tabular}]{%
  \includegraphics[width=0.24\columnwidth]{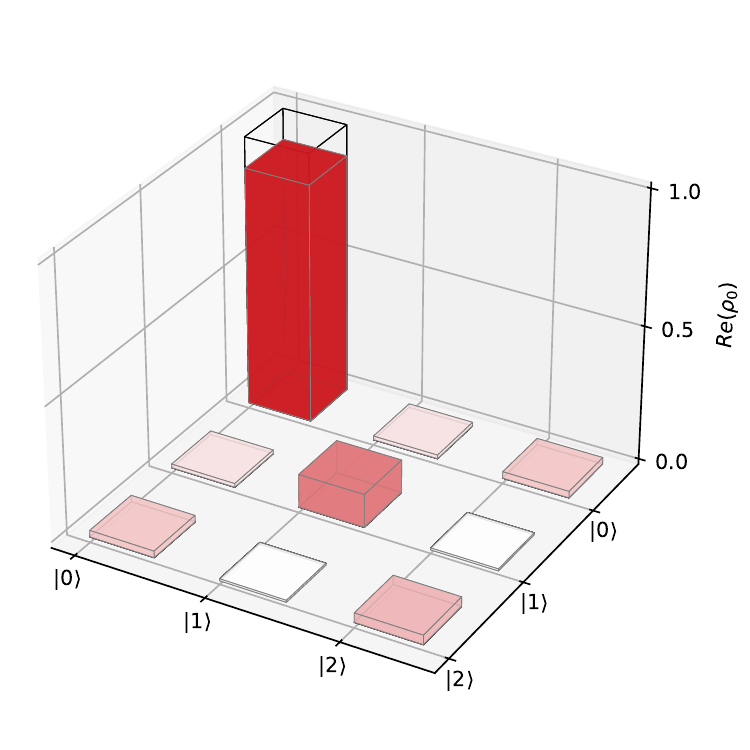}%
}\hfill
\subfloat[\label{subfig:b} \begin{tabular}{c}
       Full $\Tilde{E_0}$\\$F_{E_0}=0.9980(28)$
\end{tabular}]{%
  \includegraphics[width=0.24\columnwidth]{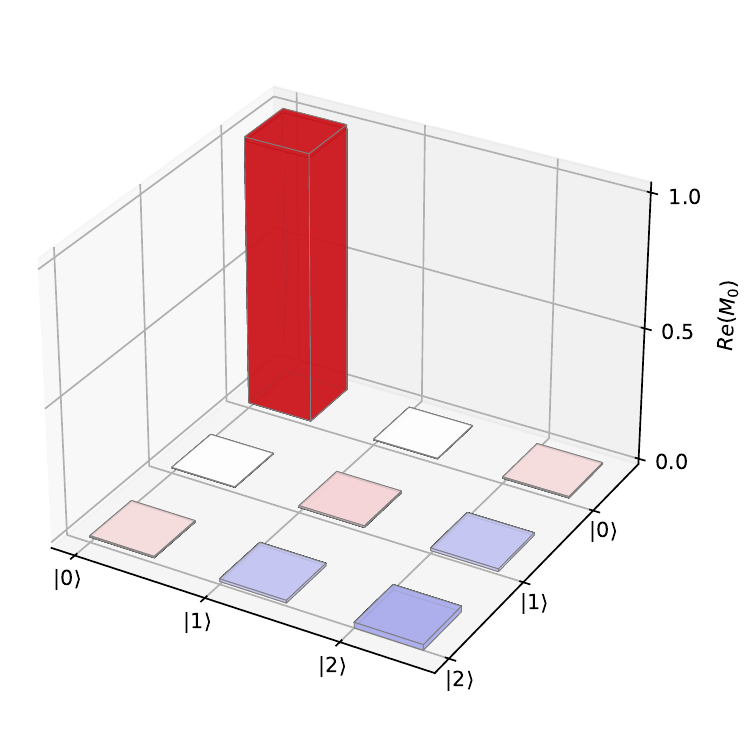}%
}
\subfloat[\label{subfig:c}\begin{tabular}{c}
       Full $\Tilde{E_1}$\\$F_{E_1}=0.9480(31)$
\end{tabular}]{%
  \includegraphics[width=0.24\columnwidth]{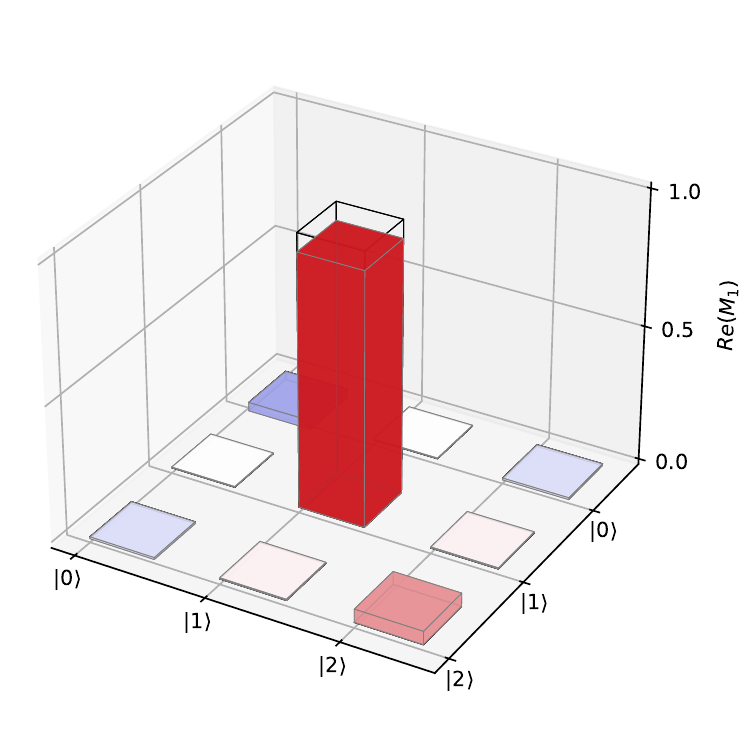}%
}\hfill
\subfloat[\label{subfig:d}\begin{tabular}{c}
       Full $\Tilde{E_2}$\\$F_{E_2}=0.9634(25)$
\end{tabular} ]{%
  \includegraphics[width=0.24\columnwidth]{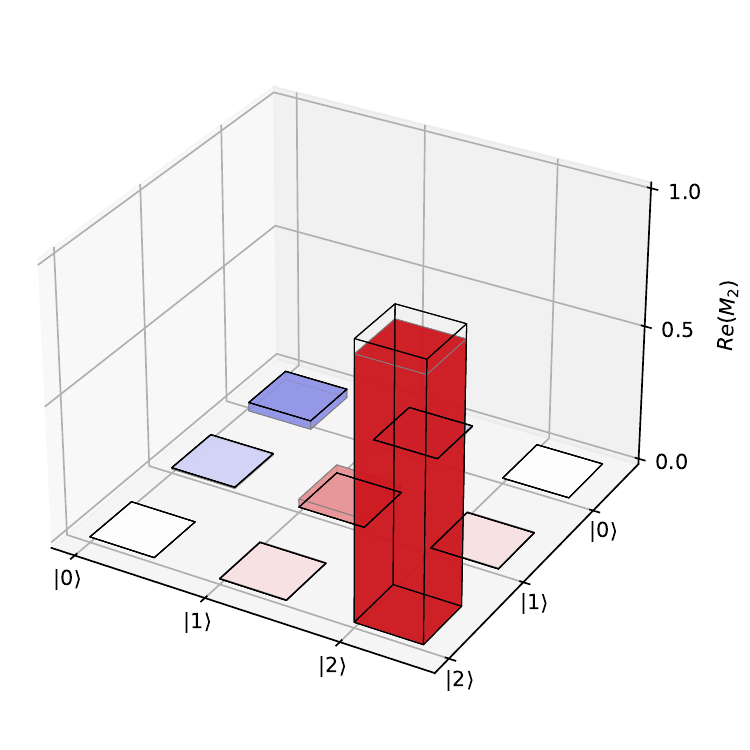}%
}\\
\subfloat[\label{subfig:e} \begin{tabular}{c}
       Static VZ $\Tilde{\rho_0}$\\$F_{\rho_0}=0.9038(36)$
\end{tabular}]{%
  \includegraphics[width=0.24\columnwidth]{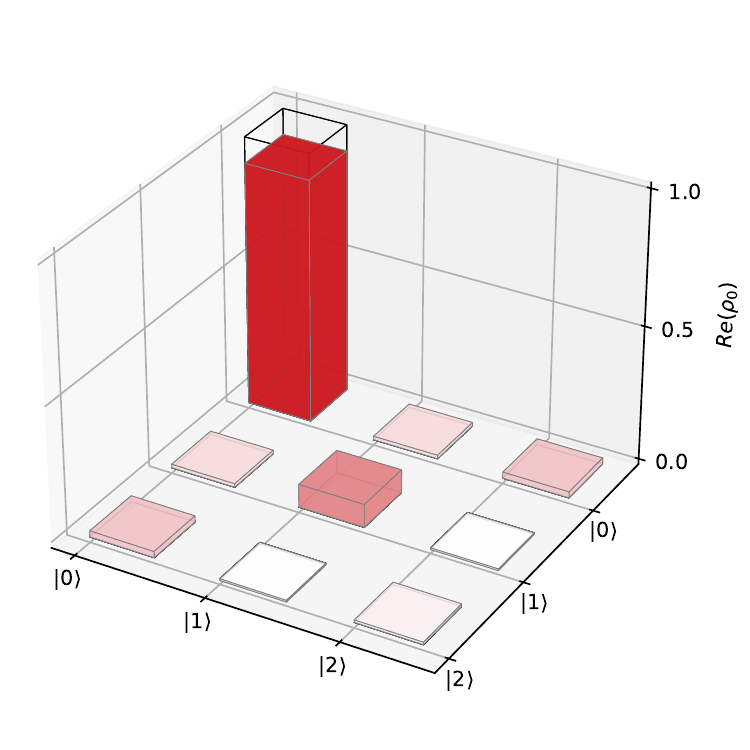}%
}\hfill
\subfloat[\label{subfig:f} \begin{tabular}{c}
       Static VZ $\Tilde{E_0}$\\$F_{E_0}=0.9845(32)$
\end{tabular}]{%
  \includegraphics[width=0.24\columnwidth]{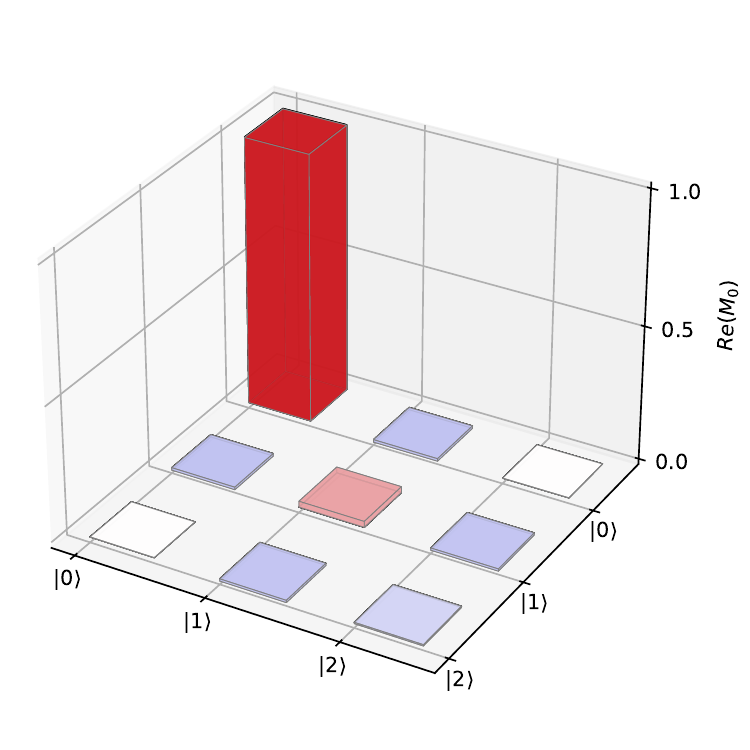}%
}
\subfloat[\label{subfig:g}\begin{tabular}{c}
       Static VZ $\Tilde{E_1}$\\$F_{E_1}=0.9326(32)$
\end{tabular}]{%
  \includegraphics[width=0.24\columnwidth]{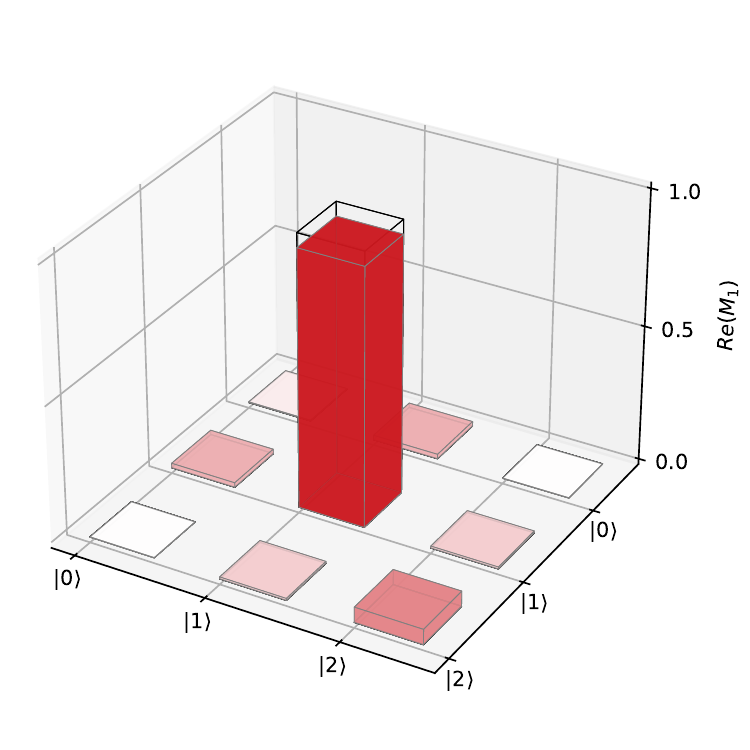}%
}\hfill
\subfloat[\label{subfig:h}\begin{tabular}{c}
       Static VZ $\Tilde{E_2}$\\$F_{E_2}=0.9490(26)$
\end{tabular} ]{%
  \includegraphics[width=0.24\columnwidth]{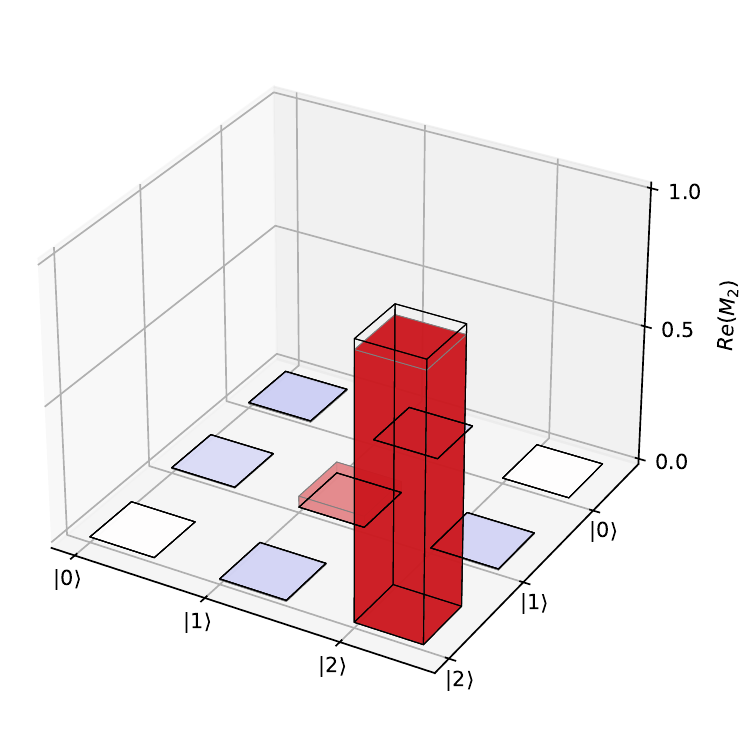}%
}
 \caption{\label{fig:rho_and_measurement_ops} Characterization of the SPAM errors:  (a) estimated initial state density matrix with the full parametrization model; (b) to (d) Estimated measurement operator $\Tilde{E}_i$  with the full parametrization model for state set to $\ket{0}$, $\ket{1}$, and $\ket{2}$, respectively. (e) estimated initial state density matrix with the model fixes virtual Z gates; (f) to (h) Estimated measurement operator $\Tilde{E}_i$  with the model fixes virtual Z gates for state set to $\ket{0}$, $\ket{1}$, and $\ket{2}$, respectively. 
 The number in brackets indicates $95\%$ confidence level.
 }
 
\end{figure}

\section{Projected error generators\label{app:error_projection}} 

The error generator $L$ describes the full error information of the reconstructed gate PTM . $L$ can be decomposed into \textit{elementry generators}, where each element has a recognizable interpretation \cite{TaxonomyErrors}. The decomposition of $L$ is given by 

\begin{equation}
\begin{aligned}
    L &= L_{\mathcal{H}} +L_{\mathcal{S}} +L_{\mathcal{C}} +L_{\mathcal{A}}\\ 
    &= \sum_P h_PH_P +\sum_P s_PS_P + \sum_{P,Q>P} c_{P,Q}C_{P,Q} + \sum_{P,Q>P} a_{P,Q}A_{P,Q}
\end{aligned}
\end{equation}
where
\begin{align}
    H_P[\rho] &= -i[P,\rho] \\
    S_P[\rho] &= P\rho P - I\rho I \\
    C_{P,Q}[\rho] &= P\rho Q + Q\rho P -\frac{1}{2}\{\{P,Q\},\rho\}\\
    A_{P,Q}[\rho] &= i(P\rho Q - Q\rho P +\frac{1}{2}\{\{P,Q\},\rho\})
\end{align}

$H_P$ is the Hamiltonian projected error generator, $S_P$ and $C_{P,Q}$ is the Pauli and Pauli-correlated projected stochastic error generator and $A_{P,Q}$ is the active projected error generator. $P$ and $Q$ are distinct bases, which are usually chosen to be Pauli matrices for qubits, and in this work we use Gellman matrices as discussed above. To further quantify the amount of coherent error and non-cohenret error, we use the metric of Jamiołkowski amplitude and Jamiołkowski probability proposed in \cite{TaxonomyErrors,Mądzik2022}. Jamiołkowski probability is a metric to quantify the incoherent error. It is defined by \cite{TaxonomyErrors}
\begin{equation}
\label{eq:Jamiołkowski_probability}
    \epsilon_J(L) = -\text{Tr}(\rho_J(L)\ket{\psi}\bra{\psi})
\end{equation}
where $\ket{\psi}$ represents a fully entangled state.

Looking at the Choi-Jamiołkowski state \cite{Choi1975CompletelyMatrices, Jamiokowski1972LinearOperators, Jiang2013Channel-stateDuality} of the error generator $\rho_{CJ}(L)$ is beneficial. Since the error generator is usually small, we can approximate $\rho_{CJ}(L) \approx \rho_{CJ}(G) - \rho_{CJ}(\hat{G}_i)$. Now $\rho_{CJ}(L)$ can be represented in a basis $\{\ket{\psi},\ket{\psi}^\prime_1,\ket{\psi}^\prime_2,...,\ket{\psi}^\prime_n\}$ where $\ket{\psi}$ is the maximum entanglement state, and $\ket{\psi}^\prime_i$ are basis orthogonal to $\ket{\psi}$. $\epsilon_J$ represents the probability of the state "jumping" from the fully entangled state to other states. Here the "jump" means the spectrum (distribution of the eigenvalues) of the density matrix changes and introduces mixed states. Therefore, $\epsilon_J$ is related to the incoherent part of the error generator. The Jamiołkowski amplitude is a metric for quantifying coherent error. It is defined by

\begin{equation}
 \begin{aligned}
    \theta_J(L) &= |(1-\ket{\psi}\bra{\psi})\rho_J(L)\ket{\psi}| \\&= \sqrt{\bra{\Psi}\rho_{CJ}(L)^2\ket{\Psi} - \bra{\Psi}\rho_{CJ}(L)\ket{\Psi}^2}
 \end{aligned}
 \end{equation}

From $\rho_{CJ}(L)$ representation, $\mathcal{E}_J$ represents the probability of the state "rotating" from the fully entangled state to other states. Here "rotates" means the spectrum of the density matrix remains the same. Therefore, it is related to the coherent part of the error generator.

\begin{algorithm}[htb]
\caption{Find Complete Measurement Basis for a Qudit \label{algo:measurement_fiducials_revised_simple}}
\SetAlgoLined
\KwIn{$d$ (Dimension of the qudit)}
\KwOut{$\textit{cliffordList}$ (List of Clifford gates)}

\textbf{Initialize:}\;
$\textit{basisList} \gets []$\;
$\textit{cliffordList} \gets []$\;
$\textit{defaultBases} \gets \text{defaultMeasurementBases()}$\; 
\ForEach{$\textit{basis} \in \textit{defaultBases}$}{
    $\text{Append}(\textit{basis}, \textit{basisList})$\;
}
$\text{Append}(\text{identityGate()}, \textit{cliffordList})$\;

\While{$len(\textit{basisList}) < d^2$}{
    \ForEach{$\text{clifford} \in \text{generateCliffordGates}(d)$}{
        $\textit{newBases} \gets \text{applyCliffordGate}(\textit{defaultBases}, \text{clifford})$\; 
        \textit{potentialAdditions} $\gets$ {list of bases from} \textit{newBases} {that are orthogonal to all in} \textit{basisList}\;
        
        \If{len(\textit{potentialAdditions}) $== d$ \text{\textbf{or}} \textit{len(potentialAdditions}) + $len(\textit{basisList}) >= d^2$ }{
            \text{Extend}(\textit{basisList}, \textit{potentialAdditions})\;
            {Append}$({clifford}, \textit{cliffordList})$\;
            break\; 
        }
    }
}
\Return{$\textit{cliffordList}$}\;
\end{algorithm}

\begin{algorithm}[htb]
\caption{Find Complete Initial State Basis for a Qudit\label{algo:preparation_fiducials}}
\SetAlgoLined
\KwIn{$d$ (Dimension of the qudit)}
\KwOut{$\textit{cliffordList}$ (List of Clifford gates)}

\textbf{Initialize:}\;
$\textit{cliffordList} \gets []$\;
$\textit{targetBasisList} \gets \text{initializeStandardBasis()}$\;
$\textit{initialState} \gets |0\rangle$\;

\While{$len(\textit{targetBasisList}) < d^2$}{
    \ForEach{$\text{clifford} \in \text{generateCliffordGates}(d)$}{
        $\textit{resultState} \gets \text{applyCliffordGate}(\textit{initialState}, \text{clifford})$\;
        \If{$\text{isOrthogonal}(\textit{newBasis}, \textit{basisList})$}{
            $\text{Append}(\textit{resultState}, \textit{targetBasisList})$\;
            $\text{Append}(\text{clifford}, \textit{cliffordList})$\;
            break\;
        }
    }
}
\Return{\textit{$cliffordList$}}\;
\end{algorithm}

\section{Selected Germs and Fiducials for GST implementation \label{app:germs}}

\begin{table}[h]
\caption{\label{tab:fiducials} Fiducials for qutrit gate set tomography.}
\centering
\begin{ruledtabular}
    \begin{tabular}{ll|ll}
    \multicolumn{2}{c}{Preparation Fiducials ($F^{(p)}$)} & \multicolumn{2}{c}{Measurement Fiducials ($F^{(m)}$)} \\\colrule
    Index & Operation & Index & Operation \\\colrule
    1 & $I$ & 1 & $I$ \\
    2 & $H.H.H.Z_2(\frac{2\pi}{3}).H$ & 2 & $Z_2(\frac{2\pi}{3}).H.H.H.Z_2(\frac{2\pi}{3}).H$ \\
    3 & $H.H.H.Z_1(\frac{2\pi}{3}).Z_2(\frac{2\pi}{3}).Z_2(\frac{2\pi}{3}).H$ & 3 & $H.H.H.Z_1(\frac{2\pi}{3}).Z_2(\frac{2\pi}{3}).H$ \\
    4 & $H.H.H.Z_1(\frac{2\pi}{3}).Z_1(\frac{2\pi}{3}).Z_2(\frac{2\pi}{3}).H$ & 4 & $H.H.H.Z_2(\frac{2\pi}{3}).H$ \\
    5 & $H.H.H.Z_1(\frac{2\pi}{3}).Z_2(\frac{2\pi}{3}).H$ &  &  \\
    6 & $H.H.H.Z_1(\frac{2\pi}{3}).Z_1(\frac{2\pi}{3}).H$ &  &  \\
    7 & $H.H.H.Z_1(\frac{2\pi}{3}).Z_1(\frac{2\pi}{3}).Z_2(\frac{2\pi}{3}).Z_2(\frac{2\pi}{3}).H$ &  &  \\
    8 & $H.H.H.Z_2(\frac{2\pi}{3}).H.Z_2(\frac{2\pi}{3})$ &  &  \\
    9 & $H.H.H.Z_1(\frac{2\pi}{3}).Z_1(\frac{2\pi}{3}).Z_2(\frac{2\pi}{3}).Z_2(\frac{2\pi}{3}).H.Z_2(\frac{2\pi}{3})$ &  &  \\
    \end{tabular}
\end{ruledtabular}
\end{table}

\begin{table}[h]
\caption{The gate set and the set of germs chosen for gate set tomography of this experiment.}
    \begin{ruledtabular}
        \begin{tabular}{ccc|ccc}
         \multicolumn{3}{c}{Gate set, $\mathcal{G}$} & \multicolumn{3}{c}{List of germs, $G_k$} \\
         \colrule
         $I$ & $H$ & $X_{01}(\frac{\pi}{2})$ & $I$ & $H$ & $X_{01}(\frac{\pi}{2})$ \\
         $X_{12}(\frac{\pi}{2})$ & $Z_1(\frac{2\pi}{3})$ & $Z_2(\frac{2\pi}{3})$ & $X_{12}(\frac{\pi}{2})$ & $Z_1(\frac{2\pi}{3})$ & $Z_2(\frac{2\pi}{3})$ \\
         & & & $X_{01}(\frac{\pi}{2})X_{12}(\frac{\pi}{2})$ & $H.Z_1(\frac{2\pi}{3})$ & $H.Z_2(\frac{2\pi}{3})$ \\
         & & & $H.X_{01}(\frac{\pi}{2})$ & $X_{12}(\frac{\pi}{2}).Z_2(\frac{2\pi}{3})$ & $X_{12}(\frac{\pi}{2}).Z_1(\frac{2\pi}{3})$ \\
         & & & $H.X_{12}(\frac{\pi}{2})$ & $X_{01}(\frac{\pi}{2}).Z_1(\frac{2\pi}{3})$ &  \\
        \end{tabular}
    \end{ruledtabular}
\end{table}

\section{Qutrit randomized benchmarking \label{app:rb}}

\para{Qutrit randomized benchmarking} We implement qutrit Clifford RB \cite{MorvanQutritBenchmarking,Emerson_2005,Unitary2designRB} to characterize the average gate fidelity and compare it with the GST result. RB fits the success probability \( P(\ket{0}) \) with the form \( A p^m + B \), where \( A \), \( B \), and \( p \) are fitting parameters, and \( m \) is the number of Clifford gates. The average error per Clifford is given by \( r = (1-p)(d-1)/d \), where \( d=2 \) for qubits and \( d=3 \) for qutrits. The average error per physical gate for qubit-like RB is \( (1-r^{1/N_g})(d-1)/d \), where \( N_g = 1.825 \) is the average number of physical gates required to implement a Clifford gate in this experiment.

\begin{figure}
    \centering
    \includegraphics[width=.6\linewidth]{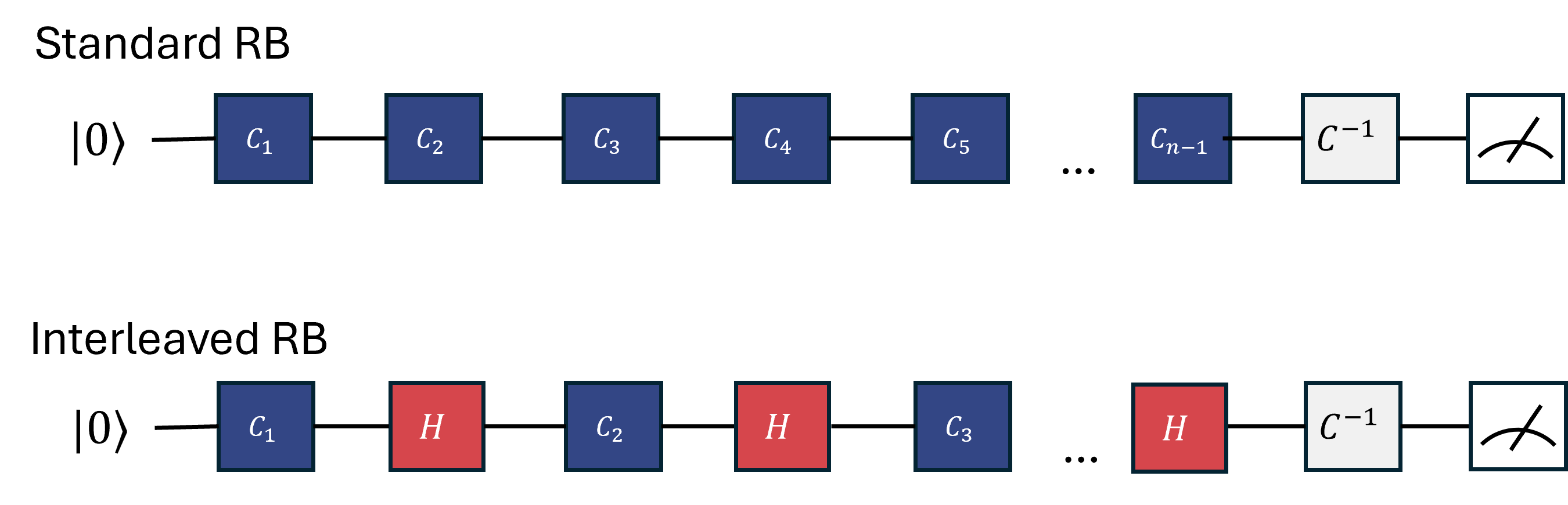}
    \caption{Experimental schemes for qutrit randomized benchmarking include standard RB and interleaved RB. In standard RB, qutrit Clifford gates are selected randomly to form a sequence. An inverse of this sequence is applied at the end to render the overall sequence as the identity. The interleaved RB experiment is utilized to extract the fidelity of the qutrit Hadamard gate in this study. It inserts the Hadamard gate after each Clifford gate in the sequence.}
    \label{fig:rb_schemes}
\end{figure}

\begin{figure}
    \centering
    \includegraphics[width=.6\linewidth]{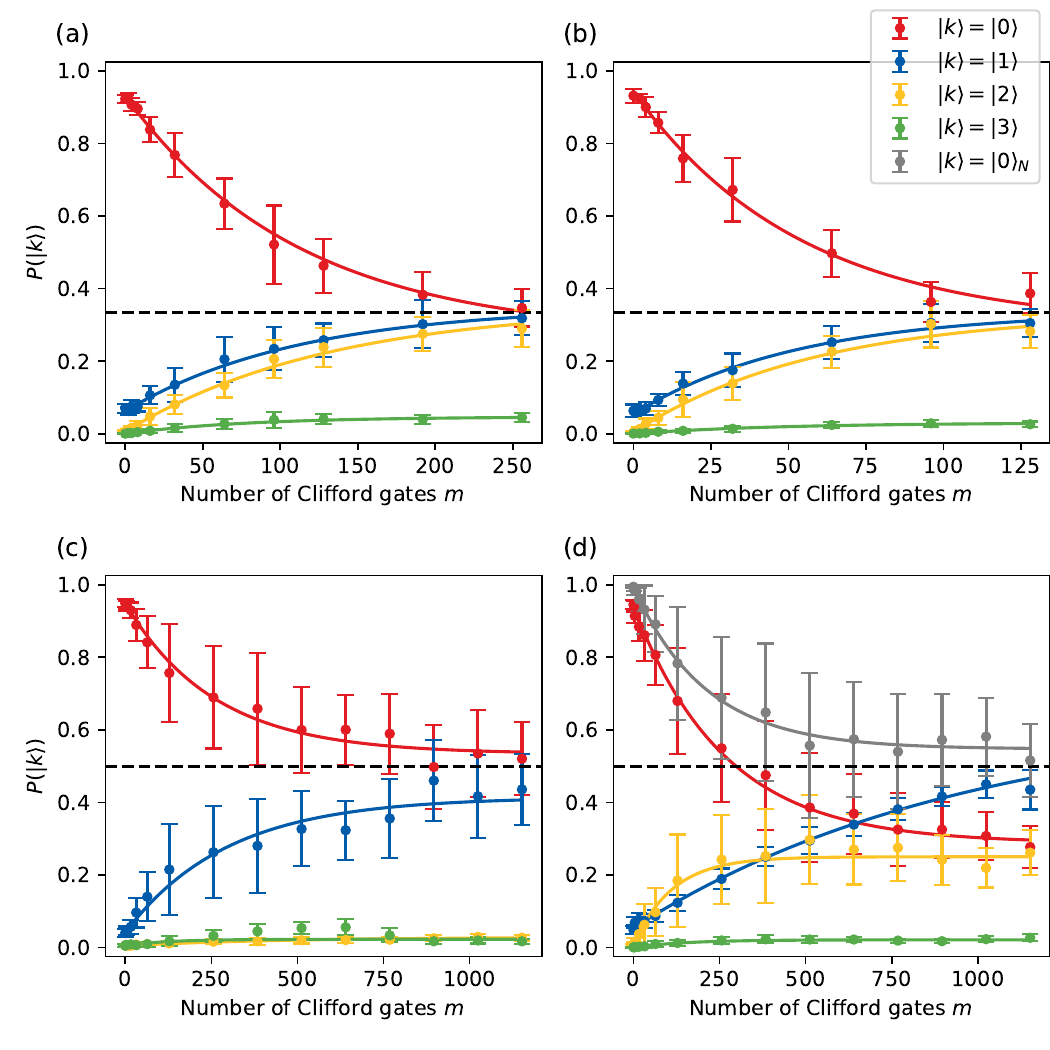}
    \caption{Randomized benchmarking (RB) results. Here, a 4-level single-shot measurement is used to distinguish four states. The measured population in the \( \ket{3} \) state is used to characterize the leakage. The error bar for the \( p \) parameter indicates one standard deviation of uncertainty. (a) Standard RB on the qutrit Clifford gate set. (b) Interleaved RB on the qutrit Hadamard gate. (c) Qubit-like RB for the \( \{\ket{0},\ket{1}\} \) subspace. (d) Qubit-like RB for the \( \{\ket{1},\ket{2}\} \) subspace. The qutrit population in \( \ket{1} \) is swapped with \( \ket{0} \) at the end of each sequence. \( \ket{0}_N \) is the renormalized population of \( \ket{0} \), where the population of the \( \ket{1} \) and \( \ket{3} \) states is excluded.}
    \label{fig:rb_parameters}
\end{figure}

The qutrit Clifford gates are synthesized with only the qutrit Hadamard gate \( H \) and the virtual Z gates. We implement single-qubit Clifford group RB on two-level subspaces \( \{\ket{0},\ket{1}\} \) and \( \{\ket{1},\ket{2}\} \). Then, we implement both standard qutrit Clifford RB and interleaved qutrit RB on the qutrit Hadamard gate~\cite{InterleavedRB,MorvanQutritBenchmarking}. The qutrit Hadamard gate fidelity can be estimated by \( r_H = (1-p_i/p)(d-1)/d \), where \( p_i \) is the exponential fitting parameter from the interleaved qutrit RB, and \( p \) is the parameter from standard qutrit RB.

We report the results from randomized benchmarking in Table \ref{tab:rb_results}. We find the average gate infidelity obtained from RB to be in reasonable agreement with our GST results; see Figure \ref{fig:Infidelity}. GST reports a lower fidelity for the \( \{\ket{0},\ket{1}\} \) subspace than for the \( \{\ket{1},\ket{2}\} \) subspace. This is likely due to the choice of local oscillator frequency being closer to the transition frequency between \( \{\ket{1},\ket{2}\} \).

\begin{table}[]
    \centering
    \caption{Infidelity obtained from RB experiments. The number in brackets in this table indicates 95\% confidence interval.}
    \label{tab:rb_results}
\begin{ruledtabular}
\begin{tabular}{ll}
Parameter            & Value $\times 10^3$  \\\hline
$\{\ket{0},\ket{1}\}$ Clifford gate infidelity &  $3.10(40) $ \\
$\{\ket{0},\ket{1}\}$ physical gate infidelity &  $1.70(22) $ \\
$\{\ket{1},\ket{2}\}$ Clifford gate infidelity  &  $2.41(26)$\\
$\{\ket{1},\ket{2}\}$ physical gate infidelity  &  $1.32(14) $ \\
Average Qutrit Clifford infidelity           & $5.93(8)$  \\
Hadamard gate infidelity & $6.3(23)$  \\
$\{\ket{0},\ket{1}\}$ leakage per qubit-like Cliffod gate  &  $0.63(19)$ \\
$\{\ket{1},\ket{2}\}$ leakage per qubit-like Clifford gate  &  $0.16(11)$ \\
Leakage per qutrit Clifford gate    & $0.669(22)$  \\
\end{tabular}
\end{ruledtabular}
\end{table}

\end{widetext}

\end{document}